\documentclass[%
 reprint,
 amsmath,amssymb,
 aps,
]{revtex4-1}

\usepackage{graphicx}
\usepackage{dcolumn}
\usepackage{bm}

\usepackage{times}
\usepackage{graphics}
\usepackage{amssymb}
\usepackage{framed}
\usepackage{braket}
\usepackage{epsfig}
\usepackage{amsmath}
\usepackage{amsfonts}

\usepackage{enumerate}

\usepackage{pifont}
\usepackage{graphicx}
\usepackage{braket}

\usepackage{array}
\usepackage{bm}
\usepackage{hyperref}
\usepackage{color}
\usepackage{makeidx}
\usepackage{mathtools}
\usepackage{bbm}
\usepackage{wrapfig}
\usepackage[skip=0pt]{caption}
\usepackage{titlesec}
\usepackage{amsmath}
\usepackage{tikz}
\usepackage{mathdots}
\usepackage{yhmath}
\usepackage{cancel}
\usepackage{color}
\usepackage{siunitx}
\usepackage{array}
\usepackage{multirow}
\usepackage{amssymb}
\usepackage{gensymb}
\usepackage{tabularx}
\usepackage{booktabs}
\usetikzlibrary{fadings}
\usepackage{hyperref}
\usetikzlibrary{cd}
\usepackage{amsthm}

\usepackage{caption}

\usepackage{adjustbox}

\usepackage[utf8]{inputenc}
\usepackage{textcomp}

\usepackage{pgfplots}
\pgfplotsset{width=8cm,compat=1.9}

\usepackage{booktabs, threeparttable}
\usepackage{siunitx}
\usepackage[linesnumbered,ruled,vlined]{algorithm2e}

\newcommand{\spn}[1]{ \text{span}\{#1 \}}

\newcommand{\C}{\mathbb{C}}

\newcommand{\pmat}[4]{\begin{pmatrix} #1 & #2 \\ #3 & #4\end{pmatrix}}

\newcommand{\sutwon}{\text{SU}(2^n)}
\newcommand{\sutwo}{\text{SU}(2)}

\newcommand{\liesutwo}{\frak{su}(2)}
\newcommand{\liesutwon}{\frak{su}(2^n)}

\newcommand{\liesun}{\frak{su}(n)}

\newcommand{\projdelta}{\text{proj}_\Delta}

\newcommand{\trace}{\text{Tr}}

\begin{document}

\preprint{APS/123-QED}

\title{Quantum Geometric Machine Learning \\for Quantum Circuits and Control}

\author{Elija Perrier}
\email{elija.t.perrier@student.uts.edu.au}
 \altaffiliation[Also at ]{eper2139@uni.sydney.edu.au; School of Economics, University of Sydney.}
\affiliation{%
 Centre for Quantum Software and Information, University of Technology, Sydney.}%

\author{Dacheng Tao}
\email{dacheng.tao@sydney.edu.au}
\affiliation{%
 UBTECH, University of Sydney}%


\author{Chris Ferrie}
\email{christopher.ferrie@uts.edu.au}
\affiliation{
 Centre for Quantum Software and Information, University of Technology, Sydney.}%

\date{\today}

\begin{abstract}
The application of machine learning techniques to solve problems in quantum control together with established geometric methods for solving optimisation problems leads naturally to an exploration of how machine learning approaches can be used to enhance geometric approaches to solving problems in quantum information processing. In this work, we review and extend the application of deep learning to quantum geometric control problems. Specifically, we demonstrate enhancements in time-optimal control in the context of quantum circuit synthesis problems by applying novel deep learning algorithms in order to approximate geodesics (and thus minimal circuits) along Lie group manifolds relevant to low-dimensional multi-qubit systems, such as SU(2), SU(4) and SU(8). We demonstrate the superior performance of greybox models, which combine traditional blackbox algorithms with prior domain knowledge of quantum mechanics, as means of learning underlying quantum circuit distributions of interest. Our results demonstrate how geometric control techniques can be used to both (a) verify the extent to which geometrically synthesised quantum circuits lie along geodesic, and thus time-optimal, routes and (b) synthesise those circuits. Our results are of interest to researchers in quantum control and quantum information theory seeking to combine machine learning and geometric techniques for time-optimal control problems.
\end{abstract}

\maketitle


\section{\label{sec:level1}Introduction}
\subsection{Overview} 
Machine learning-based approaches to solving theoretical and applied problems in quantum control have gained considerable traction over recent years as researchers leverage access to enhanced computational resources in order to solve numerical optimisation problems. Concurrently, geometric control techniques in which the tools of differential geometry and topology are applied to problems in quantum information processing have been applied in a variety of quantum control programmes \cite{Khaneja_Glaser_2001, Khaneja_2009, KhanejaReiss2005, Ekert_Ericsson_Hayden_Inamori_Jones_Oi_Vedral_2000}. The synthesis of geometry and quantum information has also recently emerged of interest to researchers in complexity geometry \cite{Brown_Susskind_2019, Lin_Susskind_2020}. It is natural therefore that the intersection between geometric and machine learning techniques in quantum control emerge as a cross-disciplinary research direction. Understanding such synergies between techniques within geometric control, quantum information processing and machine learning offers promising techniques within theoretical and applied quantum computational research, with potential application across other research domains.

In this work, we extend previous research seeking to combine techniques from geometric control, quantum information processing and machine learning in order to synthesise time-optimal quantum circuits for multi-qubit quantum systems. The development of techniques for improving the time-optimality of quantum circuit synthesis is of interest to researchers across the spectrum of theoretical \cite{FarhiQAOA} and applied quantum information science given the difficulties and challenges of synthesising quantum circuits for desired computations, let alone time-optimal ones. We approach this ubiquitous problem by extending geometric methods for generating approximate normal subRiemannian geodesic (and thus time-optimal) paths along certain Lie group manifolds of interest to quantum information processing (such as SU$(2^n)$) with tailored deep learning-based machine learning techniques. 

Our results consist of: (1) an evaluation of certain existing approaches for approximating geodesics along Lie manifolds via discrete sequences of unitary propagators; (2) determination of the optimal set of controls for generating discrete approximations to geodesic sequences of unitaries in SU$(2^n)$ for application in multi-qubit systems; and (3) demonstration of the utility of adopting so-called `greybox' machine learning architectures \cite{Youssry_Paz-Silva_Ferrie_2020} which combine `whitebox' architectures, i.e. prior information (such as known laws of quantum mechanics) with `blackbox' architectures, such as various neural network architectures, into synthesising quantum circuits.

\subsection{Problem description}
The focus of this work is on the development of novel machine learning architectures that leverage results from subRiemannian geometry in a quantum control setting. Such techniques are of relevance to practitioners within quantum control for a variety of reasons. First, as we discuss below in our explication of subRiemannian geometry in quantum control settings, subRiemannian control problems are a generalisation of standard Riemannian control problems in that they represent a more general form of Riemannian geometry. 

Second, subRiemannian quantum control problems arise where only a subset of the full Lie algebra (of generators) is itself directly accessible. This is of direct relevance to the majority of quantum control cases which may be envisioned for quantum computing devices in which one does not have access to the full set of underlying generators, say for arbitrary multi-qubit (qudit) systems with a limited gate set. Most quantum control problems are in fact, when characterised geometrically, subRiemannian quantum control problems. 

Third, is the result that synthesis of quantum circuits (i.e. sequences of unitary propagators) in a time-optimal fashion using geometric techniques (in which time-optimality is equated with generating discretised approximations to minimal distance geodesics on underlying Lie group manifolds) in fact may call for subRiemannian rather than Riemannian geometric techniques. 
The reason for this is that in order to generate such geodesic approximations, it is often beneficial (and in some cases necessary) to restrict the underlying control subalgebra of generators to a subset of the full Lie algebra. For many multi-qubit systems, quantum circuits are more likely to approximate geodesics (and thus be characterised as time optimal) where the generating Lie algebra is restricted to what are known as one- and two-body Pauli operators (tensor products of at most two standard Pauli operators), rather than the full Lie algebra. These three issues - the prevalence of subRiemmanian geometric features in quantum control problems, the restricted availability of generators when undertaking control and the need to synthesise circuits in a time-optimal fashion - motivate the use of geometric techniques applied in this work. 
\subsection{New contributions}
In this work, we report a number of experimental results based upon simulations of machine learning models for quantum circuit synthesis.

First, we report improved machine learning architectures for quantum circuit synthesis. We demonstrate in-sample improvements by standard metrics including MSE and average operator fidelity training, validation and generalisation by several orders of magnitude compared with relevant state of the art methods. We demonstrate that customised deep learning architectures which utilise a combination of standard and bespoke neural network layers, together with customised objective functions (such as fidelity measures) of relevance to quantum information processing, achieve superior results. This approach is denoted as `greybox' machine learning, is charactersied by models that combine known prior assumptions about quantum information processing with machine learning architectures. We demonstrate enhanced performance of greybox over blackbox models.

Second, we report an improvement on previous work combining subRiemannian geometric training with data and deep learning \cite{Swaddle_Noakes_Smallbone_Salter_Wang_2017} to synthesise quantum circuits. We show that optimal sets of controls may be obtained using a feed-forward fully-connected, Gated Recurrent Unit (GRU) Recurrent Neural Network (RNN) and custom geometric machine learning models. However, we also report on difficulties in usefully adapting such approaches for generalisation. 
Third, we demonstrate that machine learning protocols seeking to learn discretised geodesic approximations in $\text{SU}(2^n)$ are particularly sensitive to hyperparameter tuning, including time for application of generators and coverage of training geodesics over manifolds of interest. We show that selection of small time-steps for discretised unitary evolution will result in geodesic approximations highly proximal to the identity in $\text{SU}(2^n)$ (as was the case in \cite{Swaddle_Noakes_Smallbone_Salter_Wang_2017}), resulting in a deterioration in the ability to (in-sample and out of sample) learn geodesic approximations to target unitaries further away (by whatever relevant distance metric or norm is adopted) along the manifolds. Improving model performance to generalise beyond the proximity of the identity is shown to require small evolutionary timescales but also an increased number of segments of the geodesic approximation, though achieving the correct balance of timescale and segmentation. 

\subsection{Structure}
The structure of this work is as follows. Part \ref{sec:quantcontrolgeom} provides an overview of key quantum control concepts and literature relevant to our experiments. It examines the formulation of quantum control problems geometrically in terms of Lie groups and differential geometry. It also explores seminal expositions from Nielsen et al. in which time-optimal quantum circuit synthesis problems are framed in terms of generating approximate geodesics along relevant group manifolds. 
Part \ref{sec:subRiemsection} details the application of subRiemannian geometric theory to quantum circuit synthesis. Part \ref{sect:expdesign} lays out the design principles behind the series of experiments undertaken to develop improved machine learning architectures for quantum circuit synthesis via approximate discretised geodesics. Readers interested only in the technical details of the architectures should skip to this section. Part \ref{sect:results} details the results of the various experiments, with discussion set-out in Part \ref{sect:discussion}. Future work and directions emerging from this research are then discussed in Part \ref{sect:conclusion}. Code for the experiments may be found at  GitHub \footnote{Codebase: \url{https://github.com/eperrier/quant-geom-machine-learning}.}.

\section{\label{sec:quantcontrolgeom}Quantum control and geometry}
\subsection{Overview} 
The necessity of quantum control for various quantum information and computation programmes globally has seen the emergent application of classical geometric control and elsewhere in an effort to solve threshold problems such as how to synthesise time optimal circuits. Nearly two decades ago, developments in applied quantum control \cite{Khaneja_Brockett_Glaser_2001,Khaneja_Glaser_Brockett_2002,Khaneja_Glaser_2001} spurned the use of geometric tools to assist in solving optimisation problems in quantum information processing contexts such as applied NMR \cite{KhanejaReiss2005}. Related work also explored the use of Lie theoretic, geometric and analytic techniques for controllability of spin particles \cite{Dal2001}. Since that time, the connections between geometry and quantum control/information processing across cross-disciplinary fields, via the explication of transformations that enable problems in one field, in this case quantum control optimisation objectives (such as minimising controls for synthesis or reachable targets) into another, namely the language of differential geometry. 
Of particular note, Nielsen et al. \cite{Nielsen_Dowling_Gu_Doherty_2006} demonstrated that calculating quantum gate complexity could be framed in terms of a distance-minimisation problem in the context of Riemannian manifolds. In that work, upper and lower bounds on quantum gate complexity, relating to the optimal control cost in synthesising an arbitrary unitary $U_T \in \text{SU}(2^n)$, demonstrating the equivalence of this problem to the geometric challenge of finding minimal distances on certain Riemannian, subRiemannian and Finslerian manifolds. Subsequently, geometric techniques were utilised \cite{Dowling_Nielsen_2008, Gu_Doherty_Nielsen_2008} to find a lower bound on the minimal number of unitary gates required to exactly synthesise $U_T$, thereby specifying a lower bound on the number of gates required to implement a target unitary. 

Research across a range of quantum control \cite{DAL2019, DAL2008} and geometric circuit synthesis \cite{Gu_Doherty_Nielsen_2008, Leifer_2008,Li_Yu_Fei_2013}  has built upon results regarding the use of geometric techniques in quantum control settings. Of interest to researchers at the intersection of geometric and machine learning approaches for quantum circuit synthesis, and the focus of this work, is a technique developed in \cite{Swaddle_Noakes_Smallbone_Salter_Wang_2017, SwadThesis} that combines subRiemannian geometric techniques with deep learning in order to approximate normal subRiemannian geodesics for synthesis of time-optimal or nearly-time optimal quantum circuits. Our results present improved machine learning architectures tailored to learning such approximate geodesics.


\subsection{\label{sec:level2}Quantum control formalism}
\subsubsection{Control formulations}
The affinity between quantum control methods and geometric control and non-control methods arises from many sources within the literature. One fundamental reason is the intimate connection between Lie algebraic formulations of control problems, in classical and quantum settings, and the differential /geometric formulations of Lie theories on the other. In typical Lie theoretic approaches to quantum control problems \cite{DAL2007book} such as synthesis of quantum circuits, the quantum unitary of interest $U$ is drawn from a Lie group $G$. A feature of Lie groups is that they are mathematical structures that are at once \textit{groups} but also \textit{differentiable manifolds}, topological structures equipped with sufficient geometric and analytical structure to enable analytic machinery, such as the tools of differential geometry, to be applied to their study \cite{doCarmo_2016}.

A typical formulation of control problems in such Lie theoretic terms takes a target unitary $U_T$ to be an element of a Lie group, such as SU($2^n$), represented as a manifold. Associated with the underlying Lie group $G$ is an Lie algebra $\frak{g}$, say $\frak{su}(2^n)$, comprising the generators of the underlying Lie group of interest. Quantum control objectives can then be characterised as attempts to synthesise a target unitary propagator \cite{Khaneja_Brockett_Glaser_2001} belonging to such a Lie group $G$ via application of generators belonging to $\frak{g}$ in a controlled manner. In the simplest (noise-free) non-relativistic settings, computation is effected via evolution from $U(0)=I$ to $U_T$ according the time-dependent Schr{\"o}dinger equation: 
\begin{align}
    U(t)&=\mathcal{T}_+ \exp\left(-i\int_{0}^t H(s) ds\right).\label{eqn:scrhodtime}
\end{align}
The above formulation may also be expressed in terms (discussed in more detail below) of time dependent drift $H_d(t)$ and control $H_j(t)$ Hamiltonians:
\begin{align}
    \dot{U}(t) = -i(H_d(t) + \sum_{j=1}^m v_j H_j(t))U(t). \label{eqn:schrodunitary}
\end{align}
The drift part of the Hamiltonian represents the (directly) `uncontrollable' aspect of evolution (and is discussed in more detail below), while the control Hamiltonians represent evolution generated by those elements (generators) of the quantum system which are controllable, namely the generators of a Lie algebra of interest, such as, in the case of qubit systems, generalised Pauli operators. The terms $v_k = v_k(t)$ represent the `control' functions (discussed below) applied to specific generators $A_k \in \frak{g}$. Sometimes $H_k$ are representative of distinct generators, hence their summation, while at other times they represent different (usually linear) combinations of generators (in which case $v_k$ represents a vector of control functions), though this is mainly a stylistic choice. The time-dependence of the Hamiltonians is encoded in these time-dependent control functions as the generators themselves are not time-dependent. While often linear, the functional time dependence can and does often assume non-linear and complicated functional forms, especially in the presence of noise. Analytically solving for the form of the control function is difficult and usually intractable for higher-order qudit systems, with numerical methods usually adopted instead \cite{KhanejaReiss2005}. One of the motivations for the use of machine learning in quantum control problems is precisely their potential utility in learning a sufficient approximation of control functions needed to achieve quantum control objectives. 

It is common (as discussed below), in appropriate circumstances, to simplify the typical time-dependent Schr{\"o}dinger equation with its time-independent discretised approximation in which evolution towards a target unitary propagator $U_T$ is approximated via a sequence of successive unitaries generated by time-independent Hamiltonians $H_j$ applied at time $t_j$ for duration $\Delta t_j$: 
\begin{align}
    U(t)&=\mathcal{T}_+ \exp\left(-i \int_{0}^t H(s) ds\right)\\
    &=\lim_{N\to\infty} \prod_{j=N}^0 \exp(-iH_j(t_j)\Delta t)\\
    &\approx \prod_{j=N}^0 \exp(-iH_j(t_j)\Delta t)\\
    &=\prod_{j=N}^0 U_j = U_N...U_j...U_0
    \label{eqn:schrodind}
\end{align}
where $\Delta t_j = \Delta t = T/N$. That is, the unitary propagator at time $t$ (from the identity) is the cumulative reverse product (forward-solved cumulant) of a sequence of $U_j$ so that $U_j = U_{j-1}...U_0$. This approximation is considered appropriate where $\Delta t$ is small by comparison to total evolution time $T$ (or equivalently total energy) and is an approximation adopted in our experiments detailed below.

\begin{figure}
\includegraphics[width=\linewidth]{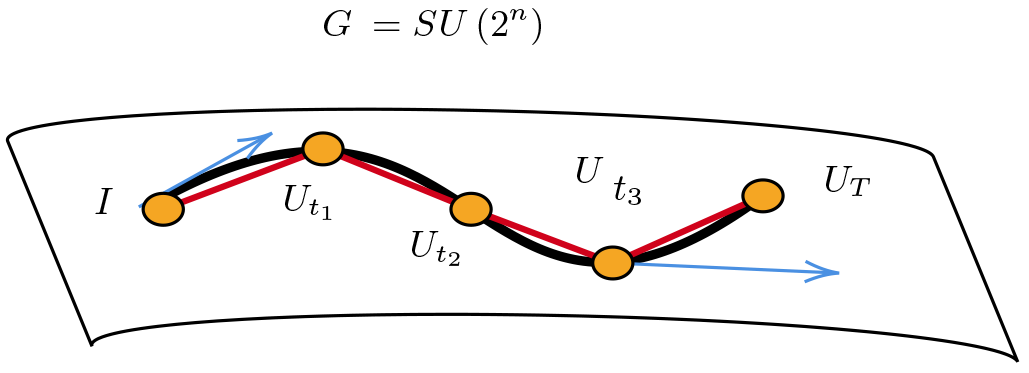}
\caption{Sketch of geodesic path. The evolution of quantum states is represented by the evolution according to Schr{\"o}dinger's equation of unitary propagators $U$ as curves (black line) on a manifold $U\in G$ generated by generators (tangent vectors) (blue) in the time-dependent case (\ref{eqn:schrodunitary}). For the time-independent case, the geodesic is approximated by evolution of discrete unitaries for time $\Delta t$, represented by red curves (shown as linear for ease of comprehension). Here $U_{t_i}$ represents the evolved unitary at time $t_i$.}
\end{figure}


Adopting this approximation allows  (\ref{eqn:schrodunitary}) to be expressed as:
\begin{equation}
    \dot{U}_j = -i(H_{d,j} + H_j)U_j. \label{eqn:control}
\end{equation}
Here, $H_{d,j} = H_d(t_j)$ designates the drift (or internal) part of the Hamiltonian at time $t_j$ and $H_j$ represents the control Hamiltonian at time $t_j$:
\begin{align}
    H_j = \sum_{k=1}^m v_k \tau_k.
\end{align}
The control Hamiltonian $H_j$, parametrised by the discretised control functions $v_k = v_k(t_j)$, are composed from (usually linear) functions of the generators $\tau_k \in \frak{g}$ (where $\dim \frak{g}=m$ and $k$ indexes the generators) belonging to the corresponding Lie algebra (such as generalised (tensor products) of Pauli operators for $SU(2^n)$. The control functions $v_k(t)$ encompass the amplitude (energy) to be applied for time $\Delta t$ (duration) over which the control Hamiltonian $H_j$ is to be applied, and typically correspond, for example, to the application of certain voltages or magnetic fields for a certain period of time. The functional form of the controls $v_k(t)$ can vary, with common (idealised) representations including Gaussian or `square' pulses.  

The objective of time optimal control is then to select the set of controls $v_j(t)$ to be applied (when using a discretised approximation) at time $t_j$ for time $\Delta t_j$ in order to synthesise $U_T$ in the shortest amount of total time. Such geometric approaches involve reparametrisation of quantum circuits, which are discrete, as approximations to geodesics on Lie group manifolds of interest to quantum information processing \cite{Nielsen_Dowling_Gu_Doherty_2006, Dowling_Nielsen_2008, Gu_Doherty_Nielsen_2008}. It is in order to solve this optimisation problem that motivates recharacterisation of problems in quantum information geometrically, such as determining and solving geodesic equations of motion.

\subsubsection{Path-length and Lie groups}
The adaptation of geometric methods and variational methods for solving optimisation problems in quantum information processing is characterised in terms of minimising distance of curves along Lie group manifolds $G$. Doing so requires selection of a metric (or cost functional) that intuitively measures the distance between elements in the associated Lie algebra which, in geometric terms, are represented by tangent vectors belonging to the associated tangent space $TG$. Cost-functionals are essentially analogous to variational functional equations such that:
\begin{align}
    C = \int_a^b g_{\alpha\beta} \frac{dx^\alpha}{dt}\frac{dx^\beta}{dt}
\end{align}
where $g_{\alpha\beta}$ represents the (not necessarily constant) metric tensor, $dx/dt$ represents the differential Lie group elements $x \in G$ with respect to the unique single-parametrisation (i.e. time) with which we are familiar. Solving the optimisation problem of interest, such as synthesising a circuit in minimal time or with minimal energy, becomes a question of minimising the cost function. 
Variational methods in this approach set $\delta C=0$ and consequently use standard techniques from variational calculus to derive respective equations of motion, differential equations whose solutions (usually) take the form of exponentiated Lie algebraic elements i.e. unitary propagators, which ultimately minimise the cost functional and solve the underlying optimisation problem.

It's worth explicating the form of cost functionals for quantum information practitioners who may be less familiar with geometric methods. In the discretised case, we essentially replace the integral with a sum over the various Hamiltonians such that we have:
\begin{align}
    C_f = \int_a^b f(H(t)) dt
\end{align}
for the continuous case and
\begin{align}
    C_f = \sum_{j=a}^b f(H(t_j))
\end{align}
where $f$ represents the control function(s) applicable to the Hamiltonian $H(t)$. By selecting the appropriate parametrisation of curves on the manifold (such as a typical parametrisation by arc-length), distance along a curve (representing evolution from one unitary, such as the identity, to another) can be equated to minimal time required to evolve (and synthesise) a target unitary $U_T$ of interest. In cases where there are multiple curves between two points, then one must select the minimal path over all such paths \cite{Nielsencomplex}.  Because minimising the cost functional depends itself upon solutions (unitaries) which are themselves generated by Lie algebraic elements subject to control functions, the optimisation problem of quantum control thus becomes a problem of identifying the optimal set (sequence) of control functions to be applied over time in order to minimise the cost functional. 

Applying standard techniques from the calculus of variations (e.g. the Pontryagin Maximum Principle \cite{Earp_Pachos_2005}) with respect to the cost functional results in the the geodesic equation of motion \cite{Nielsencomplex, Dowling_Nielsen_2008} which specifies the path that minimises the action and which is typically (for constant metric Riemannian manifolds) is given by:
\begin{equation}
    \frac{d^2x^j}{dt^2} +\Gamma_{kl}^j \frac{dx^k}{dt}\frac{dx^l}{dt}=0 \label{eqn:geodesicconst}
\end{equation}
where $x=x(t) \in G$ are the unitary group elements while $dx/dt \in \frak{g}$ represent the differential operators (tangent vectors/generators) of the associated Lie algebra. Also in (\ref{eqn:geodesicconst}) it is implied that the form of applicable geodesic $g_{\alpha\beta}$ is itself identical across the manifold (which may not always be the case). $\Gamma_{kl}^j$ represent Christoffel terms obtained by variation with respect to the metric.  Given a small arc along a geodesic on a Riemannian manifold, the remainder of the geodesic path is completely determined by the geodesic equation. Solutions to the geodesic equation are, in the continuous case curves and in the discrete case approximations to curves, on the manifold of interest. In the discrete case, such discretised curves are interpretable as quantum circuits which are time-optimal when such geodesics also represent the minimal distance curve linking two unitary group elements on a manifold. In this way, variational methods leveraging geometric techniques and characterisation may be utilised for synthesising quantum circuits.

\subsubsection{Accessible controls and drift Hamiltonians}
Minimising cost functionals in the way described above involves understanding what in classical control theory is described as the set of \textit{accessible} controls available: those unitaries which may be synthesised via application of the controls are termed reachable targets \cite{JURD}. Designing appropriate machine learning algorithms using geometric methods or otherwise thus requires information on the form of control function and generators that are available to reach a desired target, such as a target unitary or quantum state. 

For a given Lie group $G$, access to the entire set of generators $\frak{g}$ renders any element $U \in G$ reachable. In quantum control settings, access to the full Lie algebraic array of generators occasionally render the problem of unitary synthesis, i.e. the sequence of generators and control pulses, analytically or trivially obtainable using geometric means, such as Euler decompositions where $G=\text{SU}(2^n)$ \cite{Boozer_2012}. In many circumstances (such as those explored below), we are constrained or seek to synthesise target unitaries $U_T$ using only a subset of the relevant Lie algebra, a subset named the \textit{control algebra} (or \textit{control subalgebra}) $\frak{k} \subset \frak{g}$. In such cases, the full set of generators is not directly accessible. However, one may still be able to reach the target unitary of interest if the elements of $\frak{k}$ may be combined (by operation of the Lie bracket or Lie derivative, as discussed below) in order to generate the remaining generators belonging $\frak{g}$, thus providing access to the $\frak{g}$ in its entirety. We distinguish such cases by denoting the first case as a case of \textit{directly} accessible controls, while the second case represents \textit{indirectly} accessible controls. 

Returning to the quantum control paradigm (\ref{eqn:control}), the drift Hamiltonian $H_d$ represents the evolution of a quantum system which cannot be \textit{directly} controlled. It may represent a noise term or the interaction of a system with an environment in open quantum systems' formulations. Where a control subalgebra $\frak{k} \subset \frak{g}$ represents only a subset of the relevant Lie algebra, we can think of the complement $\frak{p} = \frak{k}^\bot$ (where $\frak{g} = \frak{p}\oplus \frak{k}$) as generators from which the drift term $H_d$, or at least elements of it (noting that, for example, in open quantum systems or non-unitary evolutions, generators are not necessarily Lie algebraic in character), above is composed, i.e. that $H_d \in \frak{p}$. 

The interaction between the drift $H_d$ (named due to its origins in fluid dynamics) and control $H_j$ Hamiltonians depends on the set of such accessible controls available to solve the quantum control problem of interest. The application of control Hamiltonians in this case represents, in effect, an attempt to `steer' a system evolving according to $H_d$ towards a desired target via the adjoint action of Lie group elements generated by $\frak{k}$ (see \cite{Khaneja_Brockett_Glaser_2001} for a discussion). 

Understanding the nature of relevant control algebras and the composition of drift Hamiltonians is an important consideration when designing and implementing machine learning architectures for geometric quantum control, including recent novel approaches applying machine learning for modelling and control of a reconfigurable photonic circuit \cite{Youssry_Chapman_Peruzzo_Ferrie_Tomamichel_2020} and to learn characteristics of $H_d$ via quantum feature engineering \cite{Youssry_Paz-Silva_Ferrie_2020}.
One of the motivations of the present work is to demonstrate the utility of being able to encode prior information about the relevant control subalgebra into machine learning protocols whose objective is the output of a time-optimal sequence of control pulses, a design choice that requires information about precisely what generators are accessible.

\subsubsection{Geometric optimisation}
Selecting the specific control subalgebra and set of control amplitudes in order to generate time-optimal quantum circuits is a difficult task. Solving this optimisation problem in quantum control and quantum circuit literature using geometric techniques follows two broad directions. One  approach uses symmetric space formalism and Cartan decompositions \cite{Khaneja_Glaser_2000, Khaneja_Glaser_2001, Khaneja_Brockett_Glaser_2001} to decompose the Lie algebra $\frak{g}$ associated with a given Lie group $G$ into symmetric and antisymmetric subalgebras such that $\frak{g} = \frak{k} \oplus \frak{p}$. Here $\frak{k}$ is the control subalgebra (containing accessible generators) and $\frak{p}$ is the subalgebra generating the non-directly controllable evolution of the system. If a suitable partition can be found satisfying certain Levi commutation relations (see \cite{DAL2008, Graaf2000}), then the Lie group can be decomposed into a Cartan decomposition $G = KAK$. By doing so, the problem of selecting the appropriate set of generators $\tau \in \frak{k}$ and control amplitudes is simplified (see Appendix  (\ref{A11:onetwobodyoperators}) for a discussion and \cite{Khaneja_Glaser_2001, DAL2008} in particular). A drawback of such methods as currently applied to problems in quantum control is their limited scope of application, namely that such methods apply only to limited symmetric space manifolds for which the methods were developed. Furthermore, the particular methods in \cite{Khaneja_Glaser_2001} used to determine the appropriate generators are limited in their generality.

An alternative but related method explored by Nielsen et al. in a range of papers \cite{Nielsencomplex, Nielsen_Dowling_Gu_Doherty_2006, Dowling_Nielsen_2008, Gu_Doherty_Nielsen_2008} approaches the problem of finding optimal generators and controls via modifying metrics applicable to cost functionals. In \cite{Nielsencomplex}, geometric techniques are applied to determine the minimal size circuit to exactly implement a specific $n$-qubit unitary operation combining variational and geometric techniques from Riemannian geometry, detailing a method for determining the lower bound of circuit complexity and circuit size by reference to the length of the local minimal geodesic between $U_T$ and $I$ (where length is determined via a Finsler metric on $\sutwon$). In later work \cite{Nielsen_Dowling_Gu_Doherty_2006, Dowling_Nielsen_2008, Gu_Doherty_Nielsen_2008}, particular metrics with penalty terms are chosen that add higher-weights to higher order Pauli operators in order to steer the generating set towards one- and two-body operators which are assessed as being optimal for geodesic synthesis (see Appendix (\ref{B1:Nielsen}) for a discussion). It is shown that in limiting cases applying the variational techniques and penalty metric of Nielsen et al., the optimal set of generators are one- and two-body terms \cite{WangLloyd2015}. 

Such variational and penalty metric-based approaches have their drawbacks, however: there are limited convergence guarantees due to, for example, the existence of exponentially Pauli geodesics (many unitaries have minimal Pauli geodesics of exponential length (see  \cite{Huang_2007})), reliance upon complicated boundary conditions, or the difficulty in discovering homeomorphic maps with which to deform known geodesics into other geodesic paths \cite{SwadThesis, Gu_Doherty_Nielsen_2008, WangLloyd2015}. The approach in the work of Nielsen et al. is also less general in that it assumes the entire distribution is the Lie algebra $\liesutwon$.

A common characteristic of both approaches in the case of $\text{SU}(2^n)$ is a preference for control algebras comprising only one- and two-body Pauli operators (operators that are tensor products of at most one or two Pauli operators) \cite{Nielsen_Dowling_Gu_Doherty_2006}. In \cite{Khaneja_Glaser_2000, Khaneja_Glaser_2001}, the rationale is that higher-order (more than two-body) generators introduce coupling terms which increase evolution time. In \cite{Nielsencomplex, Dowling_Nielsen_2008, Gu_Doherty_Nielsen_2008}, this rationale manifests in the imposition of penalty metrics upon higher-order terms in cost functionals. This approach penalises higher-order generators by assigning to them a higher weighting in the metric, thereby penalising higher-order terms in the cost function which seeks to minimise the metric of interest (in the case of \cite{Dowling_Nielsen_2008}, often Finslerian metrics $F$).  

Thus there are strong motivations for preferencing one- and two-body generator control subalgebras when devising strategies for quantum circuit synthesis. It can be shown that for higher-order SU$(2^n)$ systems, three or more body generators can themselves be composed via one- and two-body generators \cite{SwadThesis} when they form a bracket-generating set \cite{Montgomery_Society_Landweber_Loss_Ratiu_Stafford_2002}. These combined results motivate the selection of a (minimal) set of one- and two-body generators that can generate the entire Lie algebra $\frak{su}(2^n)$. This is a characteristic of the \textit{bracket-generating} set (or \textit{distribution}) $\Delta$ adopted in \cite{Swaddle_Noakes_Smallbone_Salter_Wang_2017}, where instead of imposing metrics or relying on decompositions to obtain optimal control subalgebras, the control subalgebras are selected initially to comprise only one- and two-body terms. Such reasoning does not guarantee the utility of one- and two-body terms per se (see \cite{SwadThesis} for technical examples) but provides a basis for potentially preferring such generators when designing optimisation protocols, such as via machine learning, to approximate geodesics.


\section{\label{sec:subRiemsection}SubRiemannian quantum circuit synthesis}
\subsection{Overview}
The difficulties of synthesising geodesics are well-known throughout geometric and control literature \cite{Noakes_1998, Frankel_2011}. The geodesically-driven control methods articulated in above face considerable challenges in terms of the complexities of the relevant boundary-value problem when adopting certain `penalty' metrics designed to enforce the geodesic constraints on Finslerian manifolds. Though analytic or numerical (including machine learning) architectures are unlikely to provide means of systematically synthesising approximate geodesics and time-optimal unitary synthesis for arbitrary propagators or higher-dimensional Lie groups, they have potential utility for lower-order qudit systems. 

In \cite{Swaddle_Noakes_Smallbone_Salter_Wang_2017, SwadThesis}, an approach leveraging subRiemannian, rather than Riemannian, geometry is adopted in order to overcome some of these barriers to quantum circuit synthesis using geodesic approximations. SubRiemannian geometry \cite{Montgomery_Society_Landweber_Loss_Ratiu_Stafford_2002, Shizume12, Nielsen_Dowling_Gu_Doherty_2006} is a generalised form of Riemannian geometry that is well-developed in classical control contexts. In its simplest description, it covers typical geometries where only a subset of the full Lie algebra $\frak{g}$ is directly accessible.

For the purposes of quantum control, it is helpful to characterise subRiemannian manifolds in simplified Lie theoretic terms (see \cite{Montgomery_Society_Landweber_Loss_Ratiu_Stafford_2002} for a more formal treatment). For a given manifold $G$, the Lie algebra $\frak{g}$ comprises generators which also form a basis of the tangent space $TG$. Curves $\gamma(t)$ along $G$ are those generated by generators $\tau \in \frak{g}$ such that the generators may be thought of as tangent vectors tangent to the curves they generate. The curves which may be generated on a manifold in many ways characterise the manifold. A distinguishing feature of Riemannian and subRiemannian manifolds is the set of accessible generators. Riemannian manifolds are characterised by full direct access to $\frak{g}$, that is, all generators in $\frak{g}$ may generate curves on $G$. In more formal language, the directions a curve may evolve along (or subalgebra of its generators) is characterised by certain subsets $\Delta$ of the tangent bundle $TG$ for a manifold $G$. The distribution is also denoted the \textit{horizontal} tangent space, which intuitively refers to tangent vectors being `tangent' and along the manifold but more formally refers to the fact that the covariant derivative of those (generating) tangent vectors $X$ along the curve is zero, that is
\begin{align*}
    \nabla_{\gamma(t)} X = 0
\end{align*}
which is characteristic of parallel transport. By contrast, it may be the case that only a subalgebra $\frak{k} \subset \frak{g}$ where $\frak{g} = \frak{k} \oplus \frak{p}$ is accessible for generation of curves on $G$. In this case, evolution of curves tangent to certain directions of tangent vectors in $\frak{p}$ is not directly possible. This set of directly inaccessible generators is orthogonal to the set $\frak{k}$ of horizontal tangent vectors and so can be thought of as in some sense \textit{vertical}. More formally, the vertical subspace of $TG$ comprises vectors $X$ whose evolution along the curve $\gamma(t)$ is such that $\nabla_{\gamma(t)} X \neq 0$ (having some component not tangent to the manifold). In this second case, the manifold is characterisable as subRiemannian rather than Riemannian. Elements of the vertical subspace $\frak{p}$ may still affect the evolution of curves, but only indirectly to the extent the generators in $\frak{p}$ are able to be generated by the application of the Lie bracket  (see (\ref{eqn:bch} below) i.e. if the distribution is bracket-generating. A number of theorems of subRiemannian geometry \cite{Montgomery_Society_Landweber_Loss_Ratiu_Stafford_2002} then guarantee the existence and uniqueness of certain normal subRiemannian geodesics on $G$ which are both unique and minimal in length. 

Thus, for generating circuits on $G=\text{SU}(2^n)$, by constructing a distribution $\Delta$ that is bracket-generating and comprising only one- and two-body generators, it can be shown \cite{SwadThesis} that normal subRiemannian geodesics may be generated which are minimal and unique, thus approximating the minimal circuits between $I$ and $U_T$. In the next section, we detail the approach in \cite{Swaddle_Noakes_Smallbone_Salter_Wang_2017} that leverages such subRiemannian geometric insights. We do so in order to provide insight into the subRiemannian machine learning detailed in Parts \ref{sec:subRiemsection} and \ref{sect:expdesign} below.

\subsection{\label{sec:level3}SubRiemannian Normal Geodesics}
The motivation behind the approach in \cite{Swaddle_Noakes_Smallbone_Salter_Wang_2017} is to solve the problem of finding time-optimal sequences of gates via approximating subRiemannian normal geodesics on Lie group manifolds \cite{Dowling_Nielsen_2008, Brandt_2010a, Brandt_2010b, Brandt_2012a, Brandt_2012b} in order to synthesise target unitary propagators $U_T \in \text{SU}(2^n)$.  The basis of the approach is to firstly adopt the time-independent approximation (\ref{eqn:schrodind}) and express $U_T$ as an approximate product of exponentials:
\begin{align}
U_T \approx U_n...U_1 \approx E(c) = \prod_j^n \underbrace{\left( \prod_k^m \exp(h c^k_j \tau_k) \right)}_{U_j} \label{eqn:swaddleunitaryseq}
\end{align}
where $U_j$ are referred to herein as (right-multiplicative or right acting) \textit{subunitaries} for convenience, again justifiable in the large $m$, small $h$ limit where $h=\Delta_j$, the evolution time of each $U_j$. The terms $c^k_j$ represent the amplitudes of the $c^k$ (square) control pulses applied to $k$ generators at time interval $t_j$ for duration $\Delta t_j = h$ to generate unitary $U_j$ (i.e. $j$ indexes the segment, $k$ indexes the control amplitude $c^k$ paired with the generators $\tau_k$). The method in \cite{Swaddle_Noakes_Smallbone_Salter_Wang_2017} in effect becomes a `bang-bang control' problem \cite{Schattler_Ledzewicz_2012} in which the time-dependent Schrodinger equation is approximated by a sequence of time-independent solutions $U_j$ where control Hamiltonians $H_j$ are applied via the application of a constant amplitude $c^j_k$ for discrete time interval $h=1/N$ (with $N$ the number of segments).
The term $E(c)$ represents an embedding function 
\begin{eqnarray}
    E: \C^{n \times m} \to \text{SU}(2^n) \\ c =(c^1_1,...,c^m_N) \mapsto \prod_j^n\prod_k^m \exp(h c^k_j \tau_k) 
\end{eqnarray}
such that $c = (c^1_1,...,c^m_n) \in \mathbbm{C}^n$ where $c_m^N = (c^1_N, c^2_N,...,c^m_N)$. Here $\tau_i$ form a basis for the bracket generating subset $\Delta \in \liesutwon$ of dimension  $m$. By comparison with the conventional control setting described above (\ref{eqn:control}), the coefficients $c^k$ would correspond to $v_j$.

Because $\Delta$ constitutes the set of generators of the entire Lie algebra $\liesutwon$ which in turn acts as the generator of its associated Lie group $\sutwon$, an arbitrary unitary $U \in \sutwon$ can be obtained to arbitrary precision with sufficiently-many products of exponentials. This results from the application of the Baker-Hausdorff-Campbell (BCH) theorem (see \cite{Nielsencomplex} for a generalised explication), namely that:
\begin{equation}
    \exp(A)\exp(B) = \exp(A + B + \frac{1}{2}[A,B]+...).
    \label{eqn:bch}
\end{equation}
The approach in \cite{Swaddle_Noakes_Smallbone_Salter_Wang_2017, SwadThesis} is to constrain application to cases where $U$ may be synthesised as a product of a polynomial in $n$ terms, meaning the number of exponentials (subunitaries) required to synthesise $U$ is at most a polynomial function of the number of sub-unitaries $n$. We discuss the effect for machine learning algorithms of increasing $n$ on outcomes such as fidelity measures below.

In the control setting discussed above (in which each $U_j$ is decomposed into its BCH product with coefficients $c^k$) each $c^k_j$ sought to be found constitutes some optimal application of the generator $\tau_k$. This is consistent with the result in \cite{Khaneja_Glaser_2001} (see Appendix (\ref{A11:onetwobodyoperators}), in which the minimum time for synthesising the target unitary propagator is given by the smallest summation of the coefficients (controls) of the generators $\sum_{i=1}^n |\alpha_i|$ which, in our notation, would be $\sum_{k=1}^m |c^k|$. 

It is worth noting that the assumption in \cite{Khaneja_Glaser_2001} and even \cite{Dowling_Nielsen_2008} and other analytic results in control is that in effect the controls can be applied `instantaneously' such that the minimum time for evolution of a unitary (via the adjoint action of control generators on drift Hamiltonians) is lower-bounded by the evolution driven by the drift Hamiltonian $H_d$. That is, many such control regimes assume that control amplitudes can be applied without energy constraints, which is equivalent to being applicable within infinitesimal time. Often this assumption is justified by the fact that a control voltage may be many orders of magnitude greater than the energy scales of the quantum systems to be controlled. In cases where control amplitudes (for example, voltages) are, in any significant sense, upper-bounded say by energy constraints, then time for optimal synthesis of circuits will of course increase as the assumption of instantaneity will not hold. For our purposes, in a bang bang control scenario and assuming evolution according to any drift Hamiltonian sets a lower-bound on evolution time, we consider the control amplitudes $c^k$ as applied for time $h$ rather than instantaneously.

\subsubsection{One- and two-body terms}

 As discussed above, the approach in \cite{Swaddle_Noakes_Smallbone_Salter_Wang_2017, SwadThesis} is to in essence circumvent the need for elaborate penalty terms in bespoke metrics to penalise higher-order generalised Pauli geodesic generators by instead simply constraining the control subalgebra, the distribution $\Delta$, to be the Kronecker product of one- and two-body Pauli operators: 
\begin{equation}
    \triangle = \text{span}\left\{\frac{i}{\sqrt{2^n}} \sigma_\iota^j, \frac{i}{\sqrt{2^n}} \sigma_\iota^k \sigma_\iota^l \right\}
\end{equation}
where $\sigma_\iota^j$ indicates the n-fold Kronecker product/tensor product of Pauli operators at position $\{1,...,j,...,m\}$ with the two-dimensional identity operator at other indices. 

The underlying approach of the geodesic approximation method in \cite{Swaddle_Noakes_Smallbone_Salter_Wang_2017, SwadThesis} is to seek to learn the inverse map: 
\begin{align}
E^{-1}: \sutwon \to \C^{m \times n} \label{eqn:inversemap}    
\end{align}
and thus, by doing so, learn the appropriate sequence of control pulses necessary to generate time optimal evolution of unitaries and, consequently, time optimal quantum circuits. The method involves generating training data in the form of normal sub-Riemannian geodesics on $\sutwon$ form $I$ to $U_T$. The exponential product (\ref{eqn:schrodind}) represents a path along the $\sutwon$ manifold, however there may be an infinity of paths between $I$ and $U_T$ such that the map $\bm{E}$ is not injective (or unique, thus minimal), meaning $\bm{E}^{-1}$ is not well-defined.

\subsection{\label{sec:level1}Generating geodesics}
To solve this uniqueness problem, \cite{Swaddle_Noakes_Smallbone_Salter_Wang_2017, SwadThesis} propose to synthesise paths that approximate minimal normal sub-Riemannian geodesics described above. To generate normal subRiemannian geodesics in $\sutwon$, \cite{Swaddle_Noakes_Smallbone_Salter_Wang_2017} limit the norm of boundary conditions (a computational efficiency choice) and apply a generalised form of the Pontryagin Maximum Principle \cite{Schattler_Ledzewicz_2012}. They follow well-established variational approaches in \cite{Sachkov_2009} where subRiemannian geodesics may be found by minimising the energy (cost) functional: 
\begin{equation}
    \mathcal{E}[\gamma] = \int_0^1 dt \langle \dot{\gamma}(t), \dot{\gamma}(t)\rangle
\end{equation}
Specifically, $\braket{\quad,\quad}$ is the restriction of the bi-invariant norm (induced by the inner product on the tangent bundle) to $\Delta \in \liesutwon$. Here the curve $\gamma(t)$ (path) varies over $t \in [0,1]$ with tangent vectors to the curve (i.e. along the vector field) given by $\dot{\gamma}(t)$. This approach uses variational methods to minimise the path length. To contextualise this formulation in Lie theoretic terms, $\gamma(t)$ represent unitaries $U(t) \in SU(2^n)$ and $\dot{\gamma}$ the corresponding tangent (Lie algebraic) vectors. Distance along a path $\gamma(t)$ generated by the tangent vectors (generators) $\dot{\gamma(t)}$ is measured in effect by metrics applied to the tangent space. The other key assumption behind this method is that the applicable metric $g_{\alpha\beta}$ is constant.

The normal subRiemannian geodesic equations arising from minimising the energy functional above can be written in differential form \cite{Sachkov_2009} as:
\begin{align}
\dot{\gamma}(t) & = u \gamma(t) \label{eqn:gammadot}\\
\dot{\Lambda} &= [\Lambda, u] \\
u &= \text{proj}_\Delta(\Lambda). \label{projsimp}
\end{align}
It is worth unpacking each of these terms in order to connect the equations above to the control and geometric formalism above and because they are integrated into the subRiemannian machine learning model detailed below. The $u$ term represents an element of the Lie algebra $u \in \Delta \subset \liesutwon$ parameterised by $t \in [0,1]$, i.e. $u: [0,1]\to \liesutwon$ with $t \mapsto u(t)$. As such, it represents the generator of evolutions on the underlying manifold $\sutwon$. For each value $t$, the curve $\gamma(t)$ represents an element of the Lie group i.e. $\sutwon$, again parametrised by $t \in [0,1]$. The $\Lambda$ terms belong also to the Lie algebra $\liesutwo$. They differ from $u$ in that while $u$ are direct elements of the distribution $\Delta$, $\Lambda(t)$ are elements of the overall Lie algebra $\liesutwon$ that are generated by the Lie-bracket between other $\Lambda$ and $u$, hence $\Lambda: [0,1] \to \liesutwon$.  

The time-derivative $\dot{\Lambda}$ refers to how the Lie bracket commutator indicates the change in a vector field along the path $\gamma(t)$. In a control setting, the Lie derivative tells us how much the generator/tangent vector $\Lambda$ changes as it is evolved along curves $\gamma(t)$ generated by elements $u$ of the control subalgebra. For parallel transport along geodesics, as mentioned above, we require this change to be such that the covariant derivative of $\Lambda_0$ as it is parallel transported along the curve is zero, that is:
\begin{align}
    \nabla_{\gamma(t)} \Lambda_0 = 0. \label{eqn:covariant}
\end{align}

The last term (\ref{projsimp}) indicates that $u$, resides in the distribution $\Delta$ by virtue of the projection of $\Lambda$ onto the distribution $\Delta$: 
\begin{equation}
    \text{proj}_\Delta(x) = \sum_i \trace( x^\dagger \tau_i) \tau_i \in \Delta. \label{eqn:projection}
\end{equation}
This projection function is important in that it ensures that the generators of $U_j$ remain within $\Delta$, facilitating the parallel transport of $\Lambda_0$ and that $U_j$ are therefore able to be synthesised from the control subalgebra in our machine learning protocols. 
Here $\gamma(t) \sim U(t)$ and $\dot{\gamma}(t) \sim \Lambda(t)$. The geodesic curves $\gamma(t)$ depend on the initial condition $\Lambda(0)$ (the `momentum' term) with the initial `position' in the manifold being the identity unitary. In the geometric control setting over Lie group manifolds, such as unitary groups, selecting an initial generalised coordinate (akin to `position' in the manifold) and generalised momentum, which in turn amounts to selecting an initial `starting' unitary from the Lie group for the evolution at $t=0$, usually the identity $U(0) \in \sutwon$  along with a starting momentum $\Lambda(0)$ drawn from the associated Lie algebra $\liesutwon$. Given these initial operators, the geodesic equations then allow determination of tuples of unitaries and generators (positions in the Lie group manifold, momenta in the Lie algebra) for any particular time  value $t \in [0,1]$. That is, they provide a formula for determining $U(t)$ and $\Lambda(t)$. The distribution (control subalgebra) determines the types of geodesics that may be evolved along. Because the distribution is bracket generating, in principle any curve along $\text{SU}(2^n)$ may be synthesised in this way (though not necessarily directly).

As noted in \cite{SwadThesis}, the above set of equations can be written as a first-order differential equation via
\begin{equation}
    \dot{\gamma}(t) = \text{proj}_\Delta(\gamma(t) \Lambda_0 \gamma(t)^\dagger)\gamma(t). \label{eqn:swadconjugation}
\end{equation}
A first-order integrator (see (\ref{eqn:firstorderintegrator})) is used to solve for $\gamma(t) = U(t)$. It is worth analysing (\ref{eqn:swadconjugation}) in light of the discussion above on conjugacy maps and their relation to time optimal geodesic paths. The $\gamma(t)$ terms in the conjugacy map:
\begin{align}
    \gamma(t) \Lambda_0 \gamma(t)^\dagger \to \Lambda_j 
\end{align}
represent the forward-solved geodesic equations \cite{Earp_Pachos_2005,Swaddle_Noakes_Smallbone_Salter_Wang_2017}. Given the initial condition $\Lambda_0$, $\gamma(t)$ here is the cumulative evolved operator in SU$(2^n)$ that is, for time-step $t_j$, we have:
\begin{align}
    \gamma(t_j) = \prod_{i=N}^j U_j
\end{align}
In this respect conjugating $\Lambda_0$ by the $\gamma(t_j)$ is equivalent to adopting a co-rotating basis or so-called moving frame for the Lie algebra (not dissimilar to how conjugation acts in a standard Euler decomposition such as in \cite{Boozer_2012}). Projecting the conjugated $\Lambda_0$ back onto the horizontal space (i.e. $\Delta$) then defines $\Lambda_j$ as $\Lambda_0$ parallel transported along the approximate geodesic. This algorithmic approximation thus achieves (a) a way to parallel transport $\Lambda_0$ and (b) a decomposition method for generating $U_j$. 
The continuous curve $\gamma(t)$ is discretised via partitioning the parametrisation interval into $N$ segments. A first-order integrator is then utilised to solve the differential equation. In continuous form, the integration equation for the unitary propagator applied over interval $\Delta t$ takes the time-dependent form (\ref{eqn:scrhodtime}):
\begin{equation}
    U(t) = \exp\left(-i \int_0^{\Delta t} \text{proj}_\Delta(\gamma(t) \Lambda_0 \gamma(t)^\dagger) dt\right)
\end{equation}
(note that the accompanying code, the imaginary unit is incorporated into $\Delta$).
In the discrete case, the curve $\gamma(t)$ is partitioned into $N$ such segments of equal parameter-interval $h=\Delta t$, indexed by $\gamma_j$ where $j = 1,...,N$. The first-order integration resolves to:
\begin{equation}
    U_j = \exp(-ih \text{proj}_\Delta(\gamma_j \Lambda_0 \gamma_j^\dagger)) \label{eqn:firstorderintegrator}
\end{equation}
where here $U_j$ are unitaries that forward-solve the geodesic equations, represented in terms of the Euler discretisation \cite{SwadThesis}:
\begin{align}
    \gamma_{j+1} &= U_j \gamma_j\\
    &=\exp(-ih \text{proj}_\Delta(\gamma_j \Lambda_0 \gamma_j^\dagger)) \gamma_j
\end{align}
where, again to reiterate, $\gamma_{j+1}$ represents the cumulative unitary propagator at time $t_{j+1}$ and $U_j$ represents the respective unitary that propagates $\gamma_j \to \gamma_{j+1}$. The Hamiltonian $H_j$ for segment $U_j$ is given by the projection onto $\Delta$:
\begin{align}
    H_j&=\text{proj}_\Delta(\gamma_j \Lambda_0 \gamma_j^\dagger) 
\end{align}
and is applied for time $h$ (though see Appendix (\ref{A:boozerswaddlecomparison}) below for nuances regarding the interpretation of $h$ and time given the imposition of $||\text{proj}_\Delta(\Lambda_0)||=||u_0||=1$). A consequence of these formal solutions is that each $H_j$ is constrained to be generated from $\Delta$. This does not mean that only unitaries directly generated by $\Delta$ are reachable, as the action of unitaries (see (\ref{eqn:bch})) gives rise to generation of generators outside $\Delta$. It is, however, of relevance to the construction of machine learning algorithms seeking to learn and reverse-engineer geodesic approximations from target unitaries $U_T$. The consequence of this requirement is that the control functions for machine learning algorithms need only model controls for generators in $\Delta$.


\section{\label{sect:expdesign}Experimental Design}
In this section, we detail our experimental design and implementation of various machine learning models that build upon and extend work in \cite{Swaddle_Noakes_Smallbone_Salter_Wang_2017} applying deep learning to the problem of approximate geodesic quantum circuit synthesis. The overall objective of our experiments was to compare the performance of variety of different machine learning architectures in simulated environments in terms of generating time-optimal quantum circuits by being trained on approximate normal subRiemannian geodesic in $\text{SU}(2^n)$ Lie groups. While other methods, such as the `shooting' method \cite{WangLloyd2015} provide alternative means of generating geodesic data,  it was shown in \cite{Swaddle_Noakes_Smallbone_Salter_Wang_2017} that such methods particularly for higher-order $\text{SU}(8)$ cases led to considerable increases in runtime compared with neural network approaches. In any case, as our primary focus in this work  was on investigating the utility of greybox approaches to geometric machine learning architectures, such alternative methods (for example, implementing the methods of \cite{Dowling_Nielsen_2008, Gu_Doherty_Nielsen_2008}) of generating geodesics or approximations thereto were not canvassed. 

\subsection{Experimental objectives}
Synthesis of geodesics for use as training data in the various machine learning protocols utilised an adapted subRiemannian approach from \cite{Swaddle_Noakes_Smallbone_Salter_Wang_2017} and \cite{Boozer_2012}. Our overall objectives required the ability to decompose a target unitary $U_T$ in order to generate the sequence $(U_j)$ from $U_T$ and, in turn, render each $U_j$ synthesisable from a set of control amplitudes applied to generators from $\Delta$. There are a variety of classical deep learning approaches that can be adopted to solving this type of supervised learning problem, including:
\begin{enumerate}
    \item Standard neural network models: such models adopt variations on simply connected or other architecture that seeks to learn an optimal configuration of hidden representations in order to output (and thus generate) the desired sequence. On their own such models tend to be \textit{blackbox} models, in which algorithms are trained to learn a mapping from inputs (training data) to outputs (labels) without any necessary interpretability or clarity about the nature of the mapping or intermediate features being generated by the network;
    \item Generative models: generative models, such as variational autoencoders (VAEs) and generative adversarial networks (GANs) seek to learn the underlying distribution of ground truth data, then use that learnt distribution to generate new similar data; and
    \item Greybox models: greybox models, as discussed further on, seek to combine domain knowledge (such as laws of physics), also known as \textit{whitebox} models, together with blackbox models into a hybrid learning protocol.
\end{enumerate}

The actual engineering, target inputs and outputs of the various machine learning models differs depending upon metrics of success and use case. For a typical quantum control problem, the sought output of the architecture is actually the sequence of control pulses $(c_j)=(c_1^k...c_2^k)$ (where $j$ indexes the relevant subunitary and $k$ the generators to generate it at segment $k$) to be implemented in order to synthesise the target unitary (i.e. apply a gate in a quantum circuit). The target unitary $U_T$ is typically one of one or more inputs into the model architecture. 

The approach in \cite{Swaddle_Noakes_Smallbone_Salter_Wang_2017} is blackbox in nature. In that case, the input to their model was (for their global decomposition algorithm) $U_T$ with label data the sequence $(U_j)$. The aim of their algorithm, a multi-layered Gated Recurrent Unit (GRU) RNN, was to learn a protocol for decomposing arbitrary $U_T \in \text{SU}(2^n)$ into the an estimated sequence $(\hat{U}_j)$ (sequences are indicated by parentheses). The individual $\hat{U}_j$ are then fed into a subsequent simple feed-forward fully-connected neural network whose output is an estimate sequence of controls $(\hat{c}_j)$ (where $c_j$ is used as a shorthand for each control amplitude $c^k$ applied to generators $\tau_k$ for segment $j$ and parentheses indicate a sequence) for generating each $\hat{U}_j$ using $\tau_k \in \Delta$. While $\hat{U}_j$ need not itself (and is unlikely to) be exactly unitary, so long as the controls $(\hat{c}_j)$ are sufficient to then input into (\ref{eqn:swaddleunitaryseq}) to generate unitary propagators, then the objective of learning the inverse mapping (\ref{eqn:inversemap}) has been achieved. No guarantees of unitarity from the learnt model are provided in \cite{Swaddle_Noakes_Smallbone_Salter_Wang_2017}, instead there is a reliance upon simply finding (\ref{eqn:inversemap}) in order to provide $(\hat{c}_j)$.
As we articulate below, while this approach in theory is feasible, in practice where unitarity is required within the network itself (as per our greybox method driven by batch fidelity objective functions), a more detailed engineering framework for the networks is required.

\subsection{Models}

\subsubsection{Geodesic deep learning architectures}
Three primary deep learning architectures were applied to the problem of learning approximations to geodesics in SU$(2^n)$: 
\begin{enumerate}
    \item a simple multi-layer feed-forward fully-connected (FC) network model implementing adaptation of \ref{eqn:swaddleunitaryseq} that learns controls $(c_j)$ trained against $(U_j)$  (the FC Greybox model);
    \item a greybox RNN model using GRU cells \cite{Lin_Yang_He_Ye_2014} in which controls $(\hat{c}_j)$ for estimated Hamiltonians $\hat{H}_j$ are learnt without being trained against (the GRU RNN Greybox model); and
    \item a fully-connected subRiemannian greybox model (the SubRiemannian model) which generates controls $(\hat{c}_j)$ by concurrently implementing (\ref{eqn:projection}) and learning the control pulses $c_{\Lambda_0}$ for the initial generator $\Lambda_0$ (that is, a model that replicates the subRiemannian normal geodesic equations while learning initial conditions for respective geodesics).
\end{enumerate}
Each model, described in more detail below, took as initial inputs the target unitary $U_T$ together with unitary sequences $(U_j)$ such that:
\begin{align*}
    U_T \approx U_n...U_1 \dot{=} (U_j).
\end{align*}
Each new model uses various neural network architectures to generate controls $(\hat{c}_j)$ for generators $\tau_k \in \Delta$ (where $H_j = \sum_k \hat{c}^k_j \tau_k)$ which are in turn evolved via customised layers implementing (\ref{eqn:schrodind}) in order to generate estimates $(\hat{U}_j)$. These estimates $(\hat{U}_j)$ were then compared using MSE loss using an operator fidelity metric against a vector of ones (as perfect fidelity will result in unity). A second metric of average operator fidelity was also adopted to provide a measure of how well on training and validation data the networks were able to synthesise $U_j$ with respect to the estimated $\hat{U}_j$.

Unlike the segmented neural networks for learning control pulses to generate specific $U_j$, the variable weights (and units) of the neural network were constructed with greater flexibility.
The models tested are set-out below which indicates the inputs, outputs and measures (here  MSE$((U_j), (\hat{U}_j))$ refers to the batch fidelity MSE described below).
\begin{center}
 \begin{tabular}{|p{2cm}|| p{1.5cm}| p{1.5cm} |p{3cm}|} 
 \hline
Model & Inputs & Outputs & Measures \\
 \hline
  FC Greybox & $U_T, (U_j)$ & $(\hat{c}_j), (\hat{U}_j), \newline (\hat{H}_j)$ & $F((U_j), (\hat{U}_j))$, MSE$((U_j), (\hat{U}_j))$  \\
 \hline
 Sub-Riemannian & $U_T, (U_j)$ & $(\hat{c}_j), (\hat{U}_j), \newline (\hat{H}_j)$ & $F((U_j), (\hat{U}_j))$, MSE$((U_j), (\hat{U}_j))$  \\
 \hline

 GRU RNN Greybox & $U_T, (U_j)$ & $(\hat{c}_j), (\hat{U}_j), \newline (\hat{H}_j)$ & $F((U_j), (\hat{U}_j))$, MSE$((U_j), (\hat{U}_j))$  \\
 \hline
 \end{tabular}
\label{table:models}
 \end{center}
For each model, the inputs to the model were the target unitary $U_T$ and its corresponding sequence of subunitaries $(U_j)$. As detailed below, the penultimate layer of each model outputs an estimated sequence of subunitaries $(\hat{U}_j)$. This estimates sequence was then compared to the true sequence $(U_j)$ using operator fidelity (see (\ref{eqn:fidelity} below). This estimate of fidelity $F((U_j), (\hat{U}_j))$ was then compared against a vector of ones (i.e. ideal fidelity) which formed the label for the models. As described below, the customised nature of the models meant intermediate outputs, including estimated control amplitude sequences $(\hat{c}_j)$, Hamiltonian estimate sequences $(\hat{H}_j)$ and $(\hat{U}_j)$ were all accessible. 

\subsubsection{Methods, training and testing procedures}
Generation of training data for each of the models tested was achieved via implementing the first-order subRiemannian geodesic equations in Python, adapting Mathematica code from \cite{Swaddle_Noakes_Smallbone_Salter_Wang_2017}. A number of adaptations and modifications to the original format of the code were undertaken: (a) where in \cite{Swaddle_Noakes_Smallbone_Salter_Wang_2017}, unitaries were parameterised only via their real components (to effect dimensionality reduction) (relying upon an analytic means of recovering imaginary components \cite{SwadThesis}), in our approach the entire unitary was realised such that $U = X + iY$. This was adopted to improve direct generation of target unitaries of interest and to facilitate fidelity calculations, such that our unitaries became expressed in terms of:
\begin{equation}
    \hat{U} = \pmat{X}{-Y}{Y}{X}
\end{equation}
where $\dim \hat{U} = \dim \text{SU}(2^{n+1})$; and (b) in certain iterations of the code for $\Lambda_0: [0,1] \to \text{SU}(2^n)$, the coefficients of the generators were derived using tanh activation functions that allowed elements of unitaries to be more accurately generated and also to test (see Appendix \ref{A:boozerswaddlecomparison}) whether the first order integrative approach did indeed generate equivalent time-optimal holonomic paths (as in \cite{Boozer_2012}).  Our greybox machine learning architecture utilised tanh functions (with a range $-1$ to $1$) rather than the range $[0,1]$. The reason for this is that by doing so, we were able to better-approximate the relevant time-optimal control functions which give rise to the generator coefficients (for example, to reproduce the holonomic paths of \cite{Boozer_2012}, one needs the coefficients to emulate the range of the sine and cosine control functions which characterise the time-optimal evolution in that case).

Furthermore, (c) one observation from \cite{Swaddle_Noakes_Smallbone_Salter_Wang_2017} was that the training data generated unitaries relatively proximal to the identity i.e. curves that did not evolve far from their origin. This is a consequence of the time interval $\Delta t$ for each generator i.e. $\Delta t = h = 1/n_{seg}$ where $n_{seg}$ is the number of segments. The consequence of this for our results was that training and validation performance was very high for $U_T$ close to the identity (that is, similar to training sets), but declined in cases for $U_T$ further away (in terms of metric distance) from the origin. This is consistent with \cite{Swaddle_Noakes_Smallbone_Salter_Wang_2017} but also consistent with the lack of generalisation performance in their model. As such, in some iterations of the experiments we scaled-up $h$ by a factor in order to seek $U_T$ which were more spread-out across the manifold. Other experiments undertaken sought to increase the extent to which training data covered manifolds by increasing the number of segments $U_j$ of the approximate geodesic while keeping $h$ fixed (between 0 and 1). We report on scale and segment number dependence of model performance below. 

In addition to these modifications, in certain experiments we also supplemented the \cite{Swaddle_Noakes_Smallbone_Salter_Wang_2017} generative code with subRiemannian training data from a Python implementation of Boozer \cite{Boozer_2012}. In this case, given the difficulty of numerically solving for arbitrary unitaries using Boozer's approach (whose solutions in the paper rely upon analytic techniques), we generated rotations about the $z$-axis by arbitrary angles $\theta$ (denoted $\eta$ in \cite{Boozer_2012}), then rotated the entire sequence of unitaries $U_j$ by a random rotation matrix. This has the effect of generating sub-Riemannian geodesics with arbitrary initial boundary conditions and rotations about arbitrary axes, which in turn provided a richer dataset for training the various neural networks and machine learning algorithms.

For $\text{SU}(2)$, the bracket-generating set $\Delta$ can be any two of the three Pauli operators. Different combinations for $\Delta$ were explored as part of our experimental process. Our experiments focused on setting our control subalgebra $\Delta = \{X,Y\}$ as this allowed ease of comparison with analytic results of \cite{Boozer_2012} and to enable assessment of how each machine learning model performed in cases where control subalgebras were limited, which was viewed as being more realistic in experimental contexts.

Test datasets for generalisation, where the trained machine learning models are tested against out of sample data, were generated using the same subRiemmanian generative code above. We also sought to test, for each of SU(2), SU(4) and SU(8), the efficacy of the models in generating sequences $(\hat{U}_j)$ that accurately evolved to randomly generated unitaries from each of those groups. The testing methodology for geodesic approximation models comprised input of the target $U_T$ of interest into the trained model with the aim of generating control pulses $(\hat{c}_j)$ from which $(\hat{U}_j)$ (and thus $\hat{U}_T$) could be generated. 

Depending on model architecture, neural network layers (either feed-forward fully-connected, RNNs or customised models) then generated variable weight $(\hat{c}_j)$. These control amplitudes are then fed into a customised Hamiltonian estimation layer which applied $(\hat{c}_j)$ to the respective generators in $\Delta$. The output of this Hamiltonian estimation layer is a sequence of control Hamiltonians $(\hat{H}_j)$ which are input into a second customised layer which implemented quantum evolution (i.e. equation (\ref{eqn:schrodind})) in order to output $(\hat{U}_j)$. A subsequent custom layer takes $(\hat{U}_j)$ and the true $(U_j)$ as inputs and calculated their fidelity i.e. it takes as inputs batches of estimates $(\hat{U}_j)$ and ground truth sequence $(U_j)$ and calculates the operator fidelity \cite{Nielsen_Chuang_2000} of each $\hat{U}_j$ and $U_j$ via:
\begin{align}
    F(\hat{U}_j,U_j) & = |\trace(\hat{U}_j^\dagger U_j)|^2/d^2
    \label{eqn:fidelity}
\end{align}
where $d = \dim U_j$. It should be noted that in this case, the unitaries are ultimately complex-valued (rather than in realised form) prior to fidelity calculations. The outputs of the fidelity layer are the ultimate output (labels) of the model (that is, the output is a batch-size length vector of fidelities). These outputs are compared to a label batch-size length vector of ones (equivalent to an objective function targeting unit fidelity). The applicable cost function used was standard MSE but applied to the difference between ideal fidelity (unity) and actual fidelity: 
\begin{align}
    C(F,1) = \frac{1}{n}\sum_{j=1}^n (1 - F(\hat{U}_j,U_j))^2
    \label{eqn:batchfidelity}
\end{align}
where here $n$ represents the chosen batch size for the models, which in most cases was 10 or a multiple thereof. It should also be noted that this approach, which we name `batch fidelity', contributed significantly to improvements in performance: previous iterations of our experiments had engineered fidelity itself as a direct loss-function using TensorFlow's low-level API, which was cumbersome,  lacking in versatility and resulting in limited improvement by comparison with batch fidelity approaches. A standard ADAM optimizer \cite{Kingma_Ba_2017} (with $\alpha=10^{-3}$) was used for all models.

\subsubsection{Geodesic architectures: Fully-connected Greybox}
To benchmark the performance of the greybox models, a blackbox model that sought to input $U_T$ and output $(\hat{c}_j)$ was constructed using a simple deep feed-forward fully-connected layer stack taking as an input $U_T$ and outputting a sequence of estimated control amplitudes $(\hat{c}_j)$. Subsequent customised layers construct estimates of Hamiltonians by applying $(\hat{c}_j)$ to the generators in $\Delta$, which are in turn used to generate subunitaries $\hat{U}_j$.

\begin{figure}[h]
\begin{center}

\includegraphics[width=\linewidth]{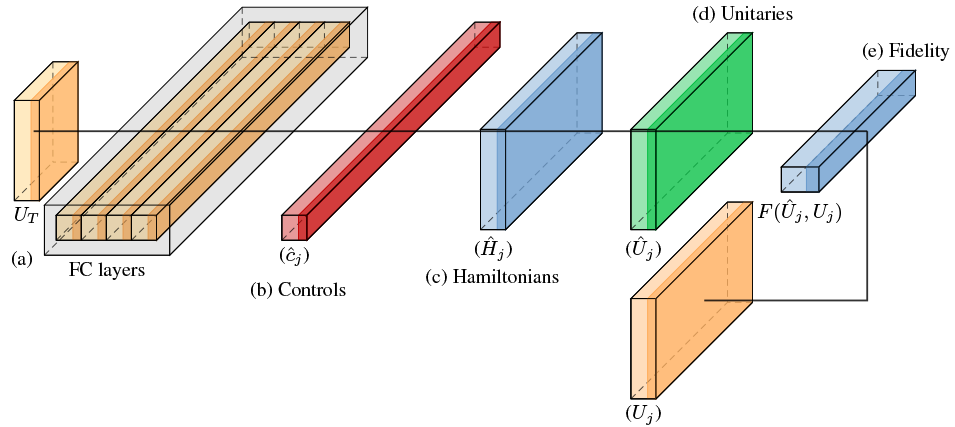}

\caption{Schema of Fully-Connected Greybox model: (a) realised $U_T$ inputs (flattened) into a stack of feed-forward fully connected layers with ReLU activations and dropout of 0.2; (b) the final dense layer in this stack outputs a sequence of controls $(\hat{c}_j)$ using using tanh activation functions; (c) these are fed into a custom Hamiltonian estimation layer produce a sequence of Hamiltonians $(\hat{H}_j)$ using $\Delta$ ; (d) these in turn are fed into a custom quantum evolution layer implementing the time-independent Schr{\"o}dinger equation to produce estimated sequences of subunitaries $(\hat{U}_j)$ which are fed into (e) a final fidelity layer for comparison with the true $(U_j)$. Intermediate outputs are accessible via submodels in TensorFlow.}
\label{diagram:fcgreybox}
\end{center}

\end{figure}

The stack comprised an initial fully-connected feed forward network with standard clipped ReLU activation functions (with dropout $\sim$ 0.2) that was fed $U_T$. This stack fed into a subsequent dense layer outputting $(\hat{c}_j)$ utilised tanh activation functions. Standard MSE loss against the label data $(c_j)$ was used (akin to the basic GRU in \cite{Swaddle_Noakes_Smallbone_Salter_Wang_2017}). The sequence $(U_j)$ was then reconstructed using $(\hat{c}_j)$ external to the model and fidelity assessed separately. A schema of the model is shown in Figure (\ref{diagram:fcgreybox}). In this variation of the feed-forward fully-connected model, a basic greybox approach that instantiated the approximation (\ref{eqn:schrodind}) was adopted. 

Greybox approaches \cite{Youssry_Chapman_Peruzzo_Ferrie_Tomamichel_2020, Youssry_Chapman_Peruzzo_Ferrie_Tomamichel_2020} represent a synthesis of `blackbox' approaches to machine learning (in which the only known data are inputs and outputs to an typical machine learning algorithm whose internal dynamics remain unknown or uninterpretable) and `whitebox' approaches, where prior knowledge of systems, such as knowledge of applicable physical laws, is engineered into algorithms. As outlined below, practically this means customising layers of neural network architecture to impose specific physical constraints and laws of quantum evolution in order to output estimated Hamiltonians and unitaries. The motivation for this approach is that it is more efficient to engineer known processes, such as the laws of quantum mechanics, into neural network architecture rather than devote computational resources to requiring the network to learn what is already known to be true (and necessary for it to function effectively) such as Schr{\"o}dinger's equation.

The greybox architecture used to estimate the control pulses necessary to synthesise each $U_j$ is set-out below. This is achieved by using $\tau_i \in \Delta$ to construct estimates of Hamiltonians $\hat{H}$ and unitaries $\hat{U}$. The inputs (training data) to the network are twofold: firstly, unitaries $\hat{U}$ generated by a Hamiltonian composed of generators in $\Delta$ with uniform randomly chosen coefficients $c^k \in [-1,1]$, where the negative values represent, intuitively, tangent vectors pointing in the opposite direction along a Lie group manifold:
\begin{align}
    \hat{H}_j &= \sum_{k=1}^{\dim|\Delta|} \hat{c}^k \tau_k \qquad \hat{c}^k \sim U[-1,1] \\
    \hat{U}_j &=  \exp(-h\hat{H}_j)
\end{align}
(recalling $i$ is absorbed into $\tau_k$ for convenience). The coefficients $c_j$  are constructed via successive feed-forward fully-connected dense layers before being applied to the generators: they are the optimal controls being sought and represent updatable weights in the network. Secondly, a tensor of training $U_j$ (generated extrinsically from $\Delta$) is separately input into the network. 

Because TensorFlow layers require output/input as real floats, $U_j$ is separated into a real $\text{Re}(U_j)$ and imaginary $\text{Im}(U_j)$ parts which are recombined in a customised layer. The specific controls being learnt by the network were accessible using standard TensorFlow techniques that allow access to intermediate output layers. This approach also provides a way to access the other intermediate outputs, such as $\hat{H}$ and $\hat{U}_j$. 

For training and validation of the model, we have the following inputs and outputs:
\begin{itemize}
    \item Inputs: $U_T$ (target unitary) and $(U_j)$ the training sequence $(U_j)$s; and
    \item Outputs: Fidelity $F(\hat{U}_j, U_j) \in [0,1]$, representing the fidelities of the estimate of the sequence $(\hat{U}_j)$ from those in the training data.
\end{itemize}

In the model, $U_T$ is fed into the feed-forward fully-connected stack followed by input into a dense flattened layer to produce a coefficient $\hat{c}^k$ for each generator $\tau_k \in \Delta$ in (\ref{eqn:swaddleunitaryseq}). Thus for a model with $n_{seg}$ segments and $\dim \Delta = d$, a total of $n_{\text{seg}} \times d$ coefficients $\hat{c}^k$ are generated. 

These are then applied to the generators $\tau_k$ in a customised Hamiltonian estimation layer. The output of this layer is then input into a unitary layer which that generates each subunitary:
\begin{align}
    \hat{U}_j = \prod_k \exp(h \hat{c}^k \tau_k)
\end{align}
in order to generate the estimated sequence of unitaries $(\hat{U}_j)$. A subsequent custom layer calculates $F(\hat{U}_j, U_j)$. The output of this layer is a (batched) vector of fidelities which are compared against a label of ones using a standard MSE loss function and Adam optimiser. This model is the simplest of the greybox models adopted in our experiments. Pseudocode for the Fully-connected Greybox model is set-out in Appendix (\ref{A0:FCG}).


\subsubsection{Geodesic architectures: GRU RNN Greybox}
The second category of deep learning architectures explored in our experiments were RNN algorithms \cite{hochreiter1997long, Goodfellow_Bengio_Courville_2016}. LSTMs are a popular deep learning tool for the modelling of sequential datasets, such as time-series or other data in which successive data depends upon preceding data points. The interested reader is directed to a number of standard texts \cite{Goodfellow_Bengio_Courville_2016} covering RNNs architecture in general for an overview. In short, RNNs are modular neural networks comprising `cells', self-enclosed neural networks consisting of inputs of training data, outputs and a secondary input from preceding cells. For sequential or time-series data, a sequence of modules are connected  for each entry or time-step in the series, $j$. The intuition behind RNNs, such as Long-Short Term Memory networks (LSTMs), is that inputs from previous time-step cells or `memories' can be carried forward throughout the network, enabling it to more accurately learn patterns in sequential non-Markovian datasets.  The original application of GRU RNNs to solving the geodesic synthesis problem was the focus of \cite{Swaddle_Noakes_Smallbone_Salter_Wang_2017}. That work utilised a relatively simple network of GRU layers  \cite{cho-etal-2014-properties}, popular due to efficiencies it can provide to training regimes. 

In the present case, the aim of the GRU RNN is to generate a model that can decompose a target unitary $U_T$ into a sequence $U_j$ reachable from $I \in \text{SU}(2^n)$. That is, the GRU RNN seeks to reverse-engineer the geodesically approximate sequence of subunitaries through a learning protocol that is itself sequential. In this model, the index $j$ of the sequence $(U_j)$ is akin to a `time slice'. At each slice $j$, the unitary $U_j$ is input into the corresponding GRU cell $G_j$ (one for each segment). The cell activation functions were set to the tanh function given its range of $[-1,1]$ accorded with the range of elements of desired subunitaries. The output of the GRU cell $G_j$ then becomes, with a certain probability, an input into the successor GRU cell $G_{j+1}$ which also takes as an input the successor subunitary $U_{j+1}$ where the function tanh over operators (matrices) is understood in the usual way (see Appendix \ref{A2:GRUNN} for background).  

The output of the GRU RNN is itself a sequence of control amplitudes $(\hat{c}_j)$ from which were then applied to generators in $\Delta$ in a custom Hamiltonian estimation layer in TensorFlow in order to construct Hamiltonian estimates and quantum evolution layers to generated estimated subunitaries $\hat{U}_j$. As with other models above, the sequence $(\hat{U}_j)$ was then input into a customised batch fidelity layer for comparison against the corresponding $(U_j)$. Our variations of the basic GRU RNN differed in that rather than simply concatenating and flattening all $(U_j)$ into a long single vector for input into a single GRU cell, each $U_j$ was associated with time-slice $j$, the objective being that, a discretised output of $(\hat{U}_j)$.

\begin{figure}[h]
\captionsetup[figure]{width=\textwidth}
\centering
\includegraphics[width=\linewidth]{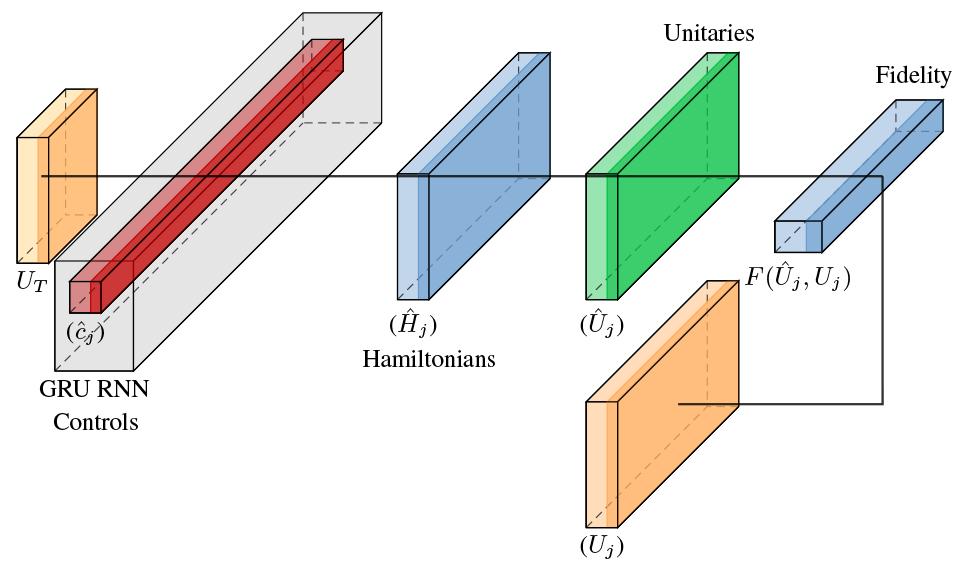}
\caption{Schema of GRU RNN Greybox model: (a) realised $U_T$ inputs (flattened) into a GRU RNN layer comprising GRU cells in which each segment $j$ plays the role of the time parameter; (b) the output of the GRU layer is a sequence of control pulses $(\hat{c}_j)$ using tanh activation functions; (c) these are fed into a custom Hamiltonian estimation layer to produce a sequence of Hamiltonians $(\hat{H}_j)$ by applying the control amplitudes to $\Delta$; (d) the Hamiltonian sequence is fed into a custom quantum evolution layer implementing the time-independent Schr{\"o}dinger equation to produce estimated sequences of subunitaries $(\hat{U}_j)$ which are fed into (e) a final fidelity layer for comparison with the true $(U_j)$. Intermediate outputs are accessible via submodels in TensorFlow.}
\label{diagram:GRU}
\end{figure}

Our main adaptation to the standard GRU RNN model was to include customised layers as described above such that the output $(\hat{U}_j)$ were themselves generated by inputting learnt coefficients $(\hat{c}_j)$ into custom Hamiltonian estimation layers (containing generators from $\Delta$), followed by quantum evolution layers (exponentiation) to generate the estimates. In this respect we followed novel approaches developed in \cite{Youssry_Chapman_Peruzzo_Ferrie_Tomamichel_2020, Youssry_Paz-Silva_Ferrie_2020}, particularly around sequential Hamiltonian estimation (though we restricted ourselves throughout to square pulse forms for $(c_j)$ only instead of also trialling Gaussian pulses). Here the aim of the GRU is to replicate the algorithmic approach in \cite{Swaddle_Noakes_Smallbone_Salter_Wang_2017}, for example learning how $\Lambda_0$ is conjugated by $U_j$ in the generation of $U_{j+1}$. Again, this represents in effect a form of `whitebox' engineering in which assured knowledge, namely how unitaries approximately evolve under the cumulative action of subunitaries, is encoded into customised layers within the network (rather than having the network `deduce' this process). Pseudocode for the GRU RNN Greybox model is set-out in Appendix (\ref{A0:GRULSTM}). A schema of the model is shown in Figure (\ref{diagram:GRU}).


\subsubsection{Geodesic architectures: SubRiemannian model}
The third model (the SubRiemannian model) architecture developed in our experiments expanded upon principles of greybox network design and subRiemannian geometry in order to generate approximations to subRiemannian geodesics. The choice of architecture was motivated by insights from the variational means of generating subRiemannian geodesics themselves \cite{Sachkov_2009, Frankel_2011}, namely that a machine learning model that effectively leveraged known or assumed knowledge regarding evolution of unitaries and their generation would perform better than more blackbox-oriented approaches. In essence the model algorithmically implemented the recursive method of generating approximate subRiemannian geodesics which relies only upon $\Lambda_0$ and $\Delta$ and relied upon learning the choice of initial condition $\Lambda_0$, rather than having to learn how to construct Hamiltonians or evolve according to the laws of quantum mechanics (which were instead dealt with via customised layers). 

The method in \cite{Swaddle_Noakes_Smallbone_Salter_Wang_2017} assumes certain prior knowledge or information in order to generate output, including: (a) the distribution $\Delta$ i.e. the control subalgebra in an experiment of interest; (b) the form of variational equations giving rise to normal subRiemannian geodesics; (c) hyperparameters, such as knowledge of the number of segments in each approximation and time-step $h$. The form of (\ref{eqn:projection}) provides (via the trace operation) the control amplitudes $\hat{c}_j$ for each generator for Hamiltonian $\hat{H}_j$. Once the initial generator $\Lambda_0$ is selected, given these prior assumptions, the output of geodesic approximations is predetermined. This characterisation was then used to design the network architecture: the inputs to the network were target unitaries $U_T$, together with the associated sequence $(U_j)$ and control subalgebra $\Delta$.

\begin{widetext}
\begin{figure}[h]
\captionsetup[figure]{width=\textwidth}
\centering
\includegraphics[width=\textwidth]{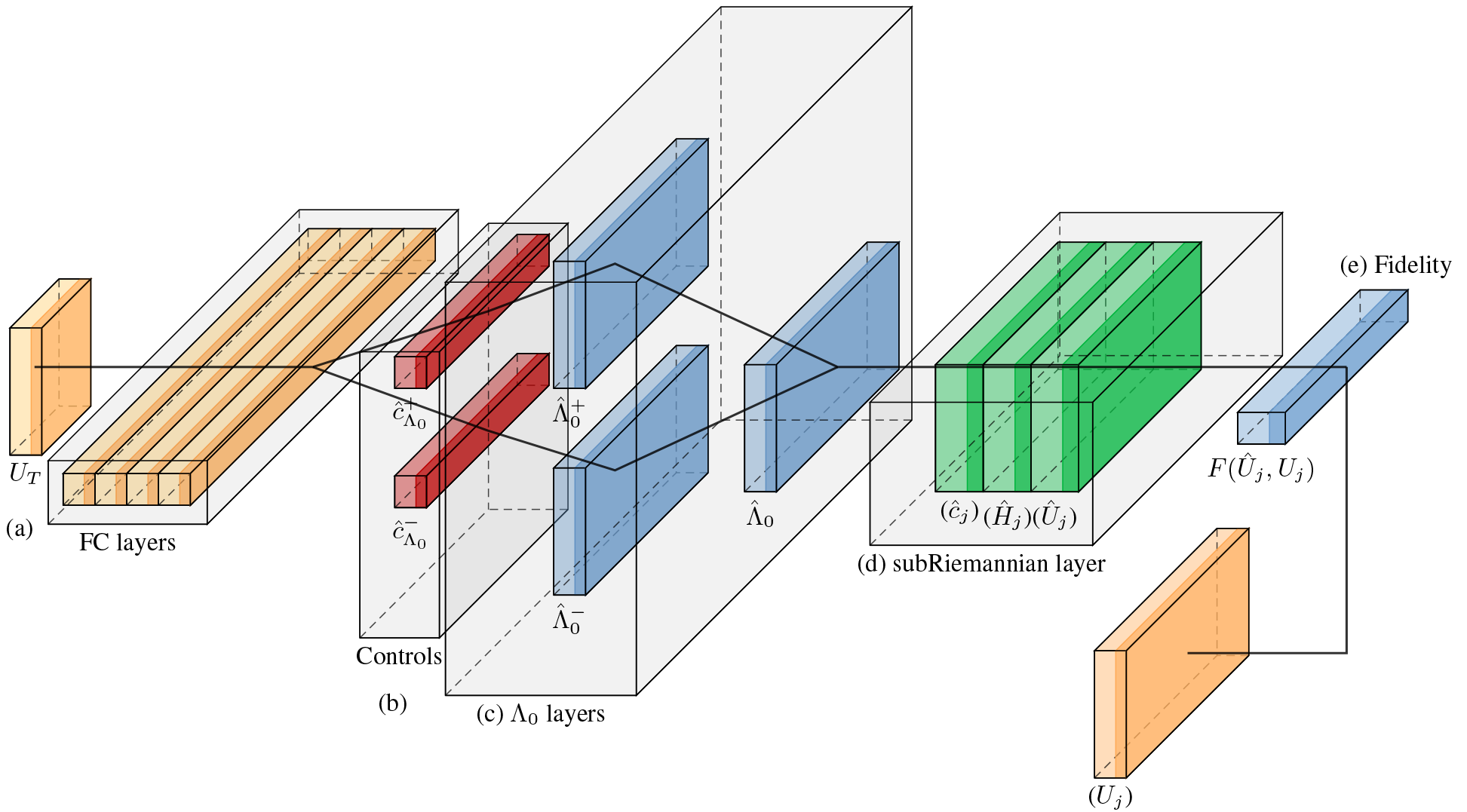}
\caption{Schema of SubRiemannian model: (a) realised $U_T$ inputs (flattened) into a set of feed-forward fully-connected dense layers (with dropout $\sim$ 0.2); (b) two layers (red) output sets of control amplitudes for estimating the positive $(\hat{c}^{+}_{\Lambda_0})$ and negative $(\hat{c}^{-}_{\Lambda_0})$ control amplitudes using tanh activation functions; (c) these are fed into two custom Hamiltonian estimation layers to produce the positive $\hat{\Lambda}^{+}_0$ and negative $\hat{\Lambda}^{-}_0$ Hamiltonians for $\Lambda_0$ using $\Delta$ or $\frak{su}(2^n)$ that are combined into a single Hamiltonian estimation $\hat{\Lambda}_0$; (d) $\hat{\Lambda}_0$ is fed into a custom subRiemannian layer which generates the control amplitudes $(\hat{c}_{j})$, Hamiltonians $(\hat{H}_{j})$ and then implements the time-independent Schr{\"o}dinger equation to produce estimated sequences of subunitaries $(\hat{U}_j)$ which are fed into (e) a final fidelity layer for comparison with the true $(U_j)$. Intermediate outputs (a) to (d) are accessible via submodels in TensorFlow. The SubRiemannian model resulted in average gate fidelity when learning representations of $(U_j)$ of over 0.99 on training and validation sets in comparison to existing GRU \& FC Blackbox models which recorded average gate fidelities of $\approx 0.70$, demonstrating the utility of greybox machine learning models in synthesising learning unitary sequences.}
\label{diagram:subR}
\end{figure}
\end{widetext}

The aim of the network was to, given the input $U_T$, learn the control amplitudes for generating the correct $\Lambda_0$ which, when input into the subRiemannian normal geodesic equations, generated the sequence $(\hat{U}_j)$ from which $U_T$ could be obtained (thus resulting in a global decomposition of $U_T$ into subunitaries evolved from the identity). Recall that $\Lambda_0$ is composed from $\frak{su}(2^n)$ or $\Delta$ depending on use case (the original paper \cite{Swaddle_Noakes_Smallbone_Salter_Wang_2017} selects $\Lambda_0 \in \frak{su}(2^n)$. This generated $\Lambda_0$ was then input into a recursive customised layer performing the projection operation (\ref{eqn:projection}) that outputs estimated Hamiltonians, followed by a quantum layer that ultimately generated the sequence $(\hat{U}_j)$. The sequence $(\hat{U}_j)$ was then input into a batch fidelity layer for comparison against the true $(U_j)$.  Once trained, the network could then be used for prediction of $\Lambda_0$, $(U_j)$, the sequence of amplitudes $(c_i)$ and $(\hat{U}_j)$, each being accessible via the creation of sub-models that access the respective intermediate custom layer used to generate such output. Pseudocode for the SubRiemannian model is set-out in Appendix (\ref{A0:SUB}). A schema of the model is shown in Figure (\ref{diagram:subR}).

As we discuss in our results section, this architecture provided among the highest-fidelity performance which is unsurprising given that it effectively reproduces the subRiemannian generative method in its entirety. One point to note is that, while this architecture generated the best performance in terms of fidelity, in terms of the actual learning protocol (i.e. the extent to which the network learns as measured by declines in loss), it was less adaptive than other architectures. That is, while having overall lower MSE, it was initialised with a lower MSE which declined less. This is not unexpected given that, in some sense, the neural network architecture combined with the whitebox subRiemannian generative procedure overdetermines the task of learning the coefficients of a single generator $\Lambda_0$ used as an initial condition. The other point to note is that in \cite{Swaddle_Noakes_Smallbone_Salter_Wang_2017}, $\Lambda_0 \in \frak{su}(2^n)$ i.e. it is drawn from the full Lie algebra, not just $\Delta$ (intuitively because it provides a random direction in the tangent space to commence evolution from). From a control perspective, however, if one only has access to $\Delta$, one cannot necessarily synthesise $\Lambda_0$, thus a second iteration of experiments where $\Lambda_0 \in \Delta$ were undertaken. The applicability of the SubRiemannian model as a means of solving the control problem is more directly related to this second case rather than the first.

\onecolumngrid


\twocolumngrid

\subsubsection{Geodesic architectures: GRU \& Fully-connected (original) model}
In order to benchmark the performance of the greybox models described above, we recreated the original global and local machine learning models utilised in \cite{Swaddle_Noakes_Smallbone_Salter_Wang_2017}. In that paper, the global model utilised a simple shallow-depth GRU RNN taking $U_T$ as inputs and outputting sequence estimates $(\hat{U}_j)$ (being trained on the true $(U_j)$). In this global decomposition, each element of each $U_j$ is in effect a trainable weight, with the GRU RNN returning the full $(\hat{U}_j)$ instead of only the control amplitudes as intermediate layers as in our GRU RNN Greybox model. The local model took $U_j$ as an input and output the coefficient control amplitude estimates $(\hat{c}_j)$ from which the sequence $(\hat{U}_j)$ could be reconstructed using $\Delta$. In \cite{Swaddle_Noakes_Smallbone_Salter_Wang_2017}, in order to reduce parameter size of the model, the original global model was trained only on the real part of $(U_j)$ on the basis that the imaginary part could be recovered via application of the unitarity constraint (see \cite{SwadThesis} for details). 

To learn the individual $Uj$ segments of the approximate geodesic unitary path, we adapted while substantially modifying the approach in \cite{Swaddle_Noakes_Smallbone_Salter_Wang_2017}. In that paper, the method of learning $Uj$ segments was adopted via feeding the real part of a vectorised (i.e. flattened) unitary $Uj$ into a simple three layer feed-forward fully connected neural network. The labels for the network were the control pulse amplitudes $c_i$.

Recreating these models it was found that using only the realised part of unitaries was insufficient for model performance overall, thus we included both real and imaginary parts both for model performance but also because it is unclear whether simply training alone on realised parts of unitaries affects the way in which the networks would integrate information about the imaginary parts. Furthermore, the approach in \cite{Swaddle_Noakes_Smallbone_Salter_Wang_2017} did not use measures such as fidelity of more utility to quantum information practitioners, thus our model extended the original models by recreating the unitaries from the estimated controls $(\hat{c}_j)$.


\section{\label{sect:results}Results}
\subsection{Overview}
The motivation behind the architectures above is to develop protocols by which time-optimal quantum circuits may be implemented via sequences of control pulses applied to quantum computational systems. In this respect, the objective is for the architectures to receive a target unitary $U_T$ as input and output a sequence of control pulses $(c_j)$ necessary to synthesise the estimate $\hat{U}_T$ that optimises fidelity $F(\hat{U}_T,U_T)$. Our experimental method sought to enable comparison of blackbox and greybox methods across the synthesis of unitary propagators (gates) in $\text{SU}(2^n)$ for $n = 1,2,3$ and higher order groups in order to achieve this objective. We also sought to gain insight into hyperparameters of model architecture by shedding light on, for example, the optimal number of segments, training examples and training data.  

Throughout our experiments, we observed that the selection of hyperparameters for both the training data and the models made a significant impact on performance. For example, as we discuss below, selection of different values for $h = \Delta t$ exhibited a noticeable impact on performance in terms of training/validation batch fidelity MSE and generalisation to test sets. For this reason, we extended our experiments to include progressively increasing values of $h$ from $h=1/n_{\text{seg}}$ to around $h\approx 1$.

Generalisation of models was tested via assessing the fidelities of $(\hat{U}_j)$ output by the trained models and also independently reconstructing $(\hat{U}_j)$ from the estimates of control coefficients $(\hat{c}_j)$. In this respect, our architecture benefited from the customised layering in that intermediate outputs of the models, such as control coefficients, sequences of estimated Hamiltonians $(\hat{H}_j)$, the actual unitary sequences $(U_j)$ and fidelities could all easily be extracted from the models using TensorFlow's standard Keras model functional API. As discussed in \cite{Youssry_Chapman_Peruzzo_Ferrie_Tomamichel_2020, Youssry_Paz-Silva_Ferrie_2020}, one of the benefits of this type of architecture is that it allows practitioners to `open' the machine learning box, as it were, to validate at intermediate steps that the whitebox outputs of the model match expectations, which in turn is useful for model tuning and engineering
\onecolumngrid
\subsection{Tables and charts}
Experimental results are set out in the tables and figures below. In Table (\ref{tab:mainresults}), each of the four models was trained and evaluated against training data from SU(2), SU(4) and SU(8). For the greybox models, batch fidelity MSE was chosen as the relevant loss function. For the GRU \& FC Blackbox model that replicated (subject to the inclusion of imaginary parts of unitaries in training) the original machine learning models in \cite{Swaddle_Noakes_Smallbone_Salter_Wang_2017}, standard MSE comparing realised unitary sequences $(U_j)$ and estimates $(\hat{U}_j)$ was used.
Average gate fidelities for training and validation data sets were also recorded, with order of magnitude of standard error provided in parentheses. Bold entries indicate the highest MSE and fidelity metrics for models trained on SU(2), SU(4) and SU(8) training data respectively.


\begin{table}[h!]
\resizebox{\textwidth}{!}{
\begin{tabular}{ |p{3cm}||p{1.35cm}|p{1.35cm}|p{1.65cm}|p{1.35cm}|p{1.35cm}|p{1.65cm}|p{1.35cm}|p{1.35cm}|p{1.65cm}| }
 \hline
 \multicolumn{10}{|c|}{Comparison table: training and validation $|$ $\Lambda_0 \in \frak{su}(2^n)$} \\
 
 \hline
 \multicolumn{1}{|c||}{Model} & \multicolumn{3}{c|}{SU(2)} & \multicolumn{3}{c|}{SU(4)} & \multicolumn{3}{c|}{SU(8)}  
 \\
 \hline
 Metric & MSE(T) & MSE(V) & Fidelity & MSE(T) & MSE(V) & Fidelity & MSE(T) & MSE(V) & Fidelity \\
    \hline
GRU \& FC Blackbox* & 3.693e-05
 & 3.559e-05
 & 0.6936(e-01) & 4.144e-05
 & 4.887e-05
 & 0.7170(e-02) &1.852e-04 & 4.447e-04 & 0.7231(e-02)  \\ 
    
    \hline
FC Greybox & 1.827e-05
 & 1.681e-05
 & 0.9964(e-05) & 3.924e-05
 & 4.156e-05
 & 0.9940(e-05) & 2.607e-04 & 2.450e-04 & 0.9842(e-05)  \\ 

\hline
SubRiemannian (XY) & \textbf{8.728e-09} & \textbf{3.211e-10}  & \textbf{0.9999(e-05)} & 1.521e-07 & 2.007e-07 & \textbf{0.9999(e-05)} & 1.024e-05 & 1.137e-04   & 0.9998(e-05)  \\

\hline
GRU RNN Greybox & 1.414e-07 & 1.348e-07 & 0.9998(e-05) & \textbf{9.019e-08}
 & \textbf{1.204e-07}
 & 0.9998(e-05) & \textbf{3.557e-06}
 & \textbf{1.186e-05}
 & \textbf{0.9998(e-05)}  \\

\hline

\end{tabular}}
\caption{Comparison table of batch fidelity MSE ($(U_j)$ and $(\hat{U}_j$)) for training (MSE(T)) and validation (MSE(V)) sets along with average operator fidelity (and order of standard deviation) for four neural networks where $\Lambda_0 \in \frak{su}(2^n)$: (a) GRU \& FC Blackbox (original) (b) FC Greybox, (c) SubRiemannian model and (d) GRU RNN Greybox model. Parameters: $h = 0.1,n_{\text{seg}}=10, n_{\text{train}}=1000$; training/validation 75/25; optimizer: Adam, $\alpha\approx$1e-3. Note*: MSE for GRU \& FC Blackbox standard MSE comparing $(U_j)$ with $\hat{U}_j$. SubRiemannian and GRU RNN Greybox models outperform blackbox models on training and validation sets with lower MSE, higher average operator fidelity and lower variance.}
\label{tab:mainresults}
\end{table}



\begin{table}[h!]
\resizebox{\textwidth}{!}{
\begin{tabular}{ |p{3cm}||p{1.35cm}|p{1.35cm}|p{1.65cm}|p{1.35cm}|p{1.35cm}|p{1.65cm}|p{1.35cm}|p{1.35cm}|p{1.65cm}| }
 \hline
 \multicolumn{10}{|c|}{Comparison table: training and validation $|$ $\Lambda_0 \in \Delta$} \\
 
 \hline
 \multicolumn{1}{|c||}{Model} & \multicolumn{3}{c|}{SU(2)} & \multicolumn{3}{c|}{SU(4)} & \multicolumn{3}{c|}{SU(8)}  
 \\
 \hline
 Metric & MSE(T) & MSE(V) & Fidelity & MSE(T) & MSE(V) & Fidelity & MSE(T) & MSE(V) & Fidelity \\
    \hline
GRU \& FC Blackbox* & 1.053e-07
 & 8.668e-08
 & 0.7180(e-02) & 1.328e-04 & 1.739e-04 & 0.9621(e-04) & 4.283e-05
 & 1.045e-04 & 0.7177(e-02)  \\

\hline
SubRiemannian (XY) & 2.616e-09
 & \textbf{9.263e-11}
  & \textbf{0.9999(e-05)} & 5.224e-08
 & 5.983e-09
 & \textbf{0.9999(e-05)} & \textbf{2.165e-07}
 & 6.874e-05
   & 0.9979(e-05)  \\

\hline
GRU RNN Greybox & \textbf{7.290e-10}
 & 7.086e-10
 & \textbf{0.9999(e-05)} & \textbf{3.478e-09}
 & \textbf{5.505e-09}
 & \textbf{0.9999(e-05)} & 2.817e-07
 & \textbf{1.092e-06}
 & \textbf{0.9994(e-05)}  \\

\hline

\end{tabular}}

\caption{Comparison table of batch fidelity MSE ($(U_j)$ v. $(\hat{U}_j$)) for training (MSE(T)) and validation (MSE(V)) sets along with average operator fidelity (and order of standard deviation) for models where $\Lambda_0 \in \Delta$: (a) GRU \& FC Blackbox (original) (b) SubRiemannian model and (c) GRU RNN Greybox model. Parameters: $h = 0.1,n_{seg}=10, n_{train}=1000$; training/validation 75/25; optimizer: Adam, $\alpha\approx$1e-3. Note*: MSE for GRU \& FC Blackbox standard MSE comparing $(U_j)$ with $\hat{U}_j$. For this case, overall the GRU RNN Greybox model performed slightly better than the SubRiemannian model, with both outperforming the GRU \& FC Blackbox model. The FC Greybox model was not tested given its inferior performance overall.}.
\label{tab:mainresultsdelta}
\end{table}


\twocolumngrid

\begin{figure}[h]
\captionsetup[figure]{width=\textwidth}
\centering
\includegraphics[width=\linewidth]{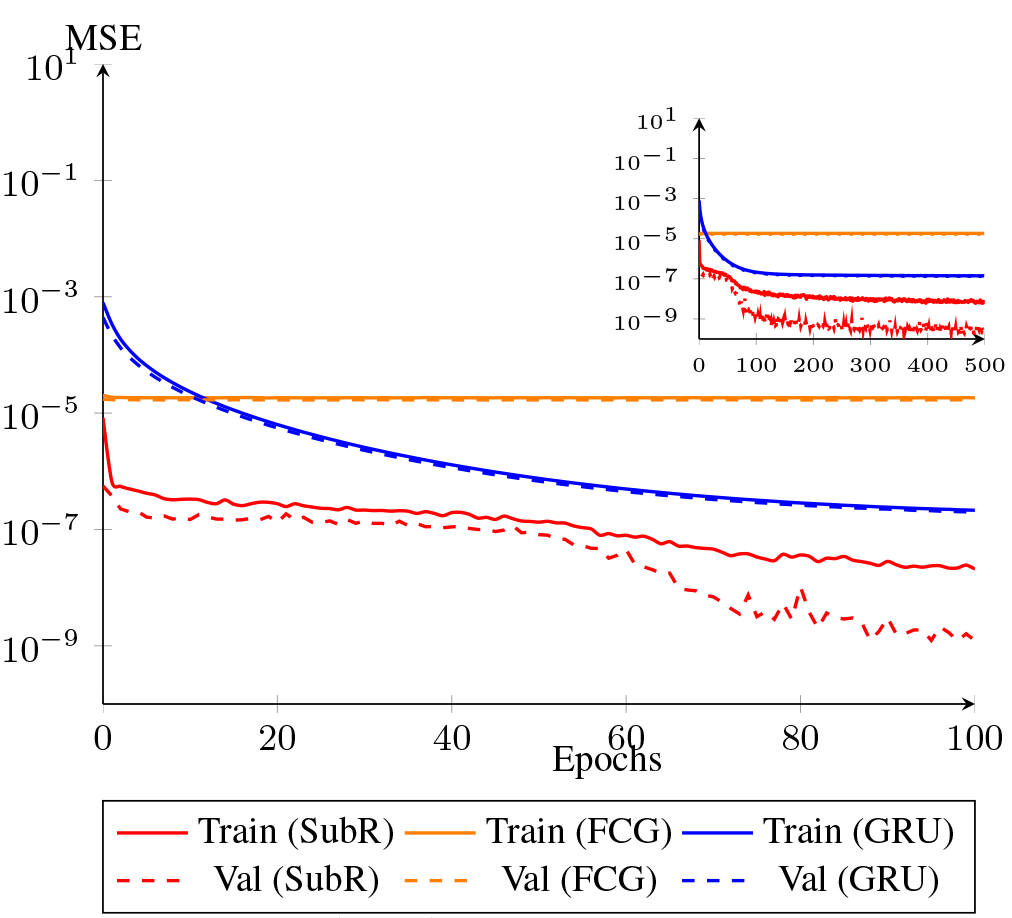}
\caption{Training and validation loss (MSE). Comparison of FC Greybox, SubRiemannian and GRU RNN Greybox models: $G=\text{SU}(2),n_{\text{train}}=1000, n_{\text{seg}}=10, h=0.1$, epochs$=500$ (first 100 shown), $\Lambda_0 \in \frak{su}(2^n)$. (Main plot - first 100 epochs) The SubRiemannian model (red) outperforms other models in terms of MSE on both training (red) and validation (dashed red) sets with MSE of order $10^{-11}$ and average operator fidelity of $0.9999$, though as can be seen by the variability in the dashed red line, exhibits more overfitting as epochs increase towards 500 (inset). The GRU RNN Greybox model (blue) exhibits smoother loss curves with comparable MSE and fidelity results. The Fully-connected Greybox (FCG) model (orange) saturates early without significant improvement. All greybox models render high average operator fidelity $> 0.99$.}
  \label{fig:losssu2threemodels}
\end{figure}


\begin{figure}[h]
\captionsetup[figure]{width=\textwidth}
\centering
\includegraphics[width=\linewidth]{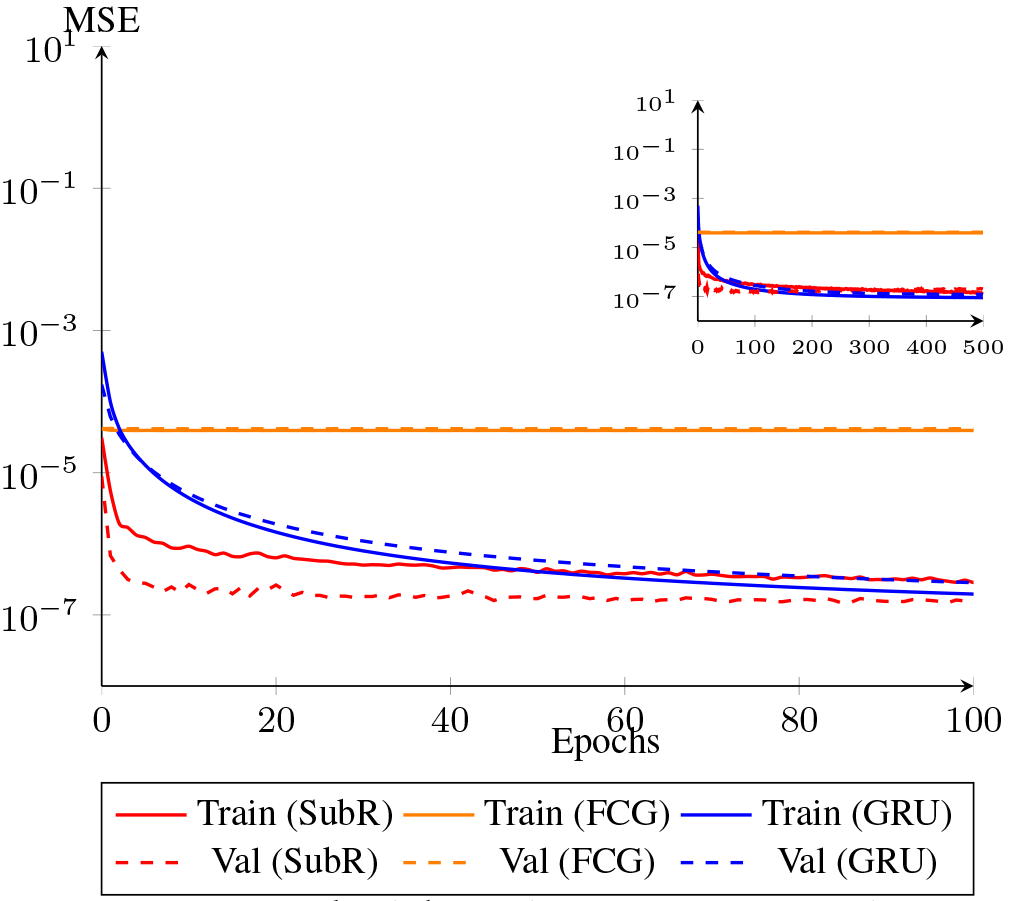}
\caption{Training and validation loss (MSE). (Main plot - first 100 epochs) Comparison of SubRiemannian, FC Greybox and GRU RNN Greybox models: $G=\text{SU}(4),n_{\text{train}}=1000, n_{\text{seg}}=10, h=0.1$, epochs$=500$ $\Lambda_0 \in \frak{su}(2^n)$. (Main plot) For the $U \in \text{SU}(4)$, the SubRiemannian model initially renders best batch fidelity MSE ($\sim 10^{-8})$ improving more quickly by comparison with other models. However, after around the 100 epoch mark, the GRU RNN Greybox performs better in terms of batch fidelity MSE (see inset for up to 500 epochs). The FC Greybox model rapidly saturates for large $n_{\text{train}}$. All models render high average operator fidelity $> 0.99$.}
  \label{fig:losssu4threemodels}
\end{figure}


\begin{figure}[h]
\captionsetup[figure]{width=\textwidth}
\centering
\includegraphics[width=\linewidth]{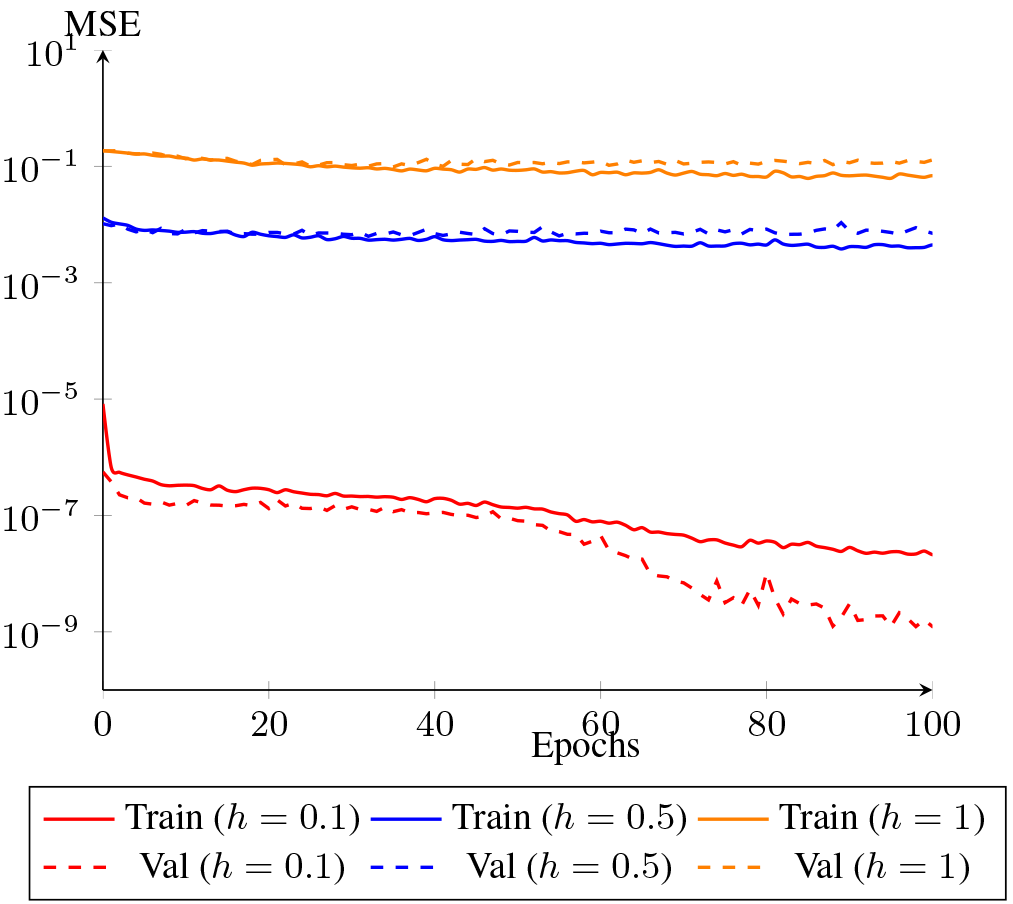}
\caption{Training and validation loss (MSE). Comparison of MSE at different time intervals. $h=0.1,0.5$ and $1$ $G=\text{SU}(2),n_{\text{train}}=1000, n_{\text{seg}}=10$, epochs$=500$, $\Lambda_0 \in \frak{su}(2^n)$: This plot shows the differences in MSE on training and validation sets as the time-step $h=\Delta t$ varies from $0.1$ to $1$. As can be seen, larger $h$ leads to deterioration in performance (higher MSE). However, smaller $h$ can lead to insufficiently long geodesics, leading to a deterioration in generalisation. Setting $h=0.1$ (red curves) exhibits the best overall performance. Even a smaller jump up to $h=0.5$ (blue curves) exhibits an increase in MSE and decrease in performance by several orders of magnitude (and similarly for $h=1$).}
  \label{fig:lossscaledependent}
\end{figure}


\twocolumngrid
\newpage
\section{\label{sect:discussion} Discussion}

\subsection{Geodesic approximation performance}
As can be seen from Table (\ref{tab:mainresults}), the in-sample (training/validation) performance of the models varied considerably between blackbox and greybox approaches. From a use-case and training data perspective, as can be seen from Table (\ref{tab:mainresults}), while the SubRiemannian and GRU RNN Greybox models outperformed the existing benchmark in \cite{Swaddle_Noakes_Smallbone_Salter_Wang_2017} in terms of in-sample batch fidelity MSE loss, we see a decline in estimations of $U_j \in \text{SU}(2^n)$ for higher $n$. MSE overall increases with dimension $n$, which is not unexpected.  

\subsection{Greybox improvements}
As can be seen from Table \ref{tab:mainresults}, the greybox models in general significantly outperformed (with fidelities around the 0.99 mark) the generic blackbox models (with fidelities in the order of 0.70) for in-sample training and validation experiments for all greybox models and all training data sets $(\Lambda \in \frak{su}(2^n)$ and $\Lambda \in \Delta$). By comparison with existing approaches in \cite{Swaddle_Noakes_Smallbone_Salter_Wang_2017} and blackbox models that seek to directly synthesise control sequences $(c_j)$ or unitary sequences $(U_j)$, the SubRiemannian and GRU RNN Greybox models outperformed (batch fidelity MSE) the FC Greybox model by several orders of magnitude. This is evident most apparently in Figures (\ref{fig:losssu2threemodels}) and (\ref{fig:losssu4threemodels}). In Figure (\ref{fig:losssu2threemodels}), representing training of the models on SU(2) data, the SubRiemannian model performs the best out of each model, though exhibits overfitting at around the 80 epoch level. These improvements were also accompanied by functional benefits such as the guarantees of unitarity of $U_j$. 


\begin{figure}[h]
\captionsetup[figure]{width=\textwidth}
\centering
\includegraphics[width=\linewidth]{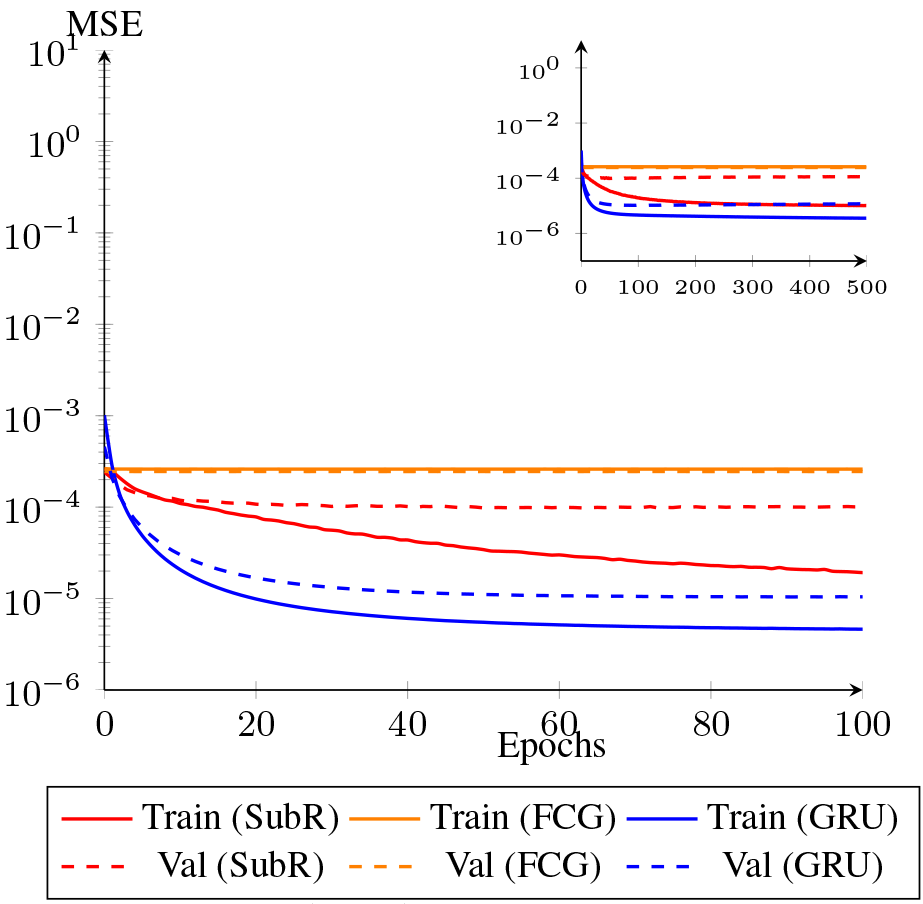}
\caption{Training and validation loss (MSE). Comparison of SubRiemannian, FC Greybox and GRU RNN Greybox models. $G=\text{SU}(8),n_{\text{train}}=1000, n_{\text{seg}}=10, h=0.1$, epochs$=500$, $\Lambda_0 \in \frak{su}(2^n)$: For $U \in \text{SU}(8)$, we see (main plot - first 100 epochs) that the GRU RNN Greybox (blue line) performs best in terms of batch fidelity MSE on training and validation sets. As shown in the inset, the GRU RNN Greybox levels out (saturates) after about 100 epochs and overall perfomed the best of each of the models and rendered average operator fidelties of around $0.998$. The SubRiemannian model (red) performed less-well than the GRU RNN, still recording high average operator fidelity but exhibiting overfitting as can be seen by the divergence of the validation (dashed) curve from the training (smooth) curve. The FC Greybox rapidly saturates for large $n_{\text{train}}$ and exhibits little in the way of learning. All models render high average operator fidelity $> 0.99$ and saturate after around 150 epochs (see inset).}
  \label{fig:su8multin1000nseg10}
\end{figure}

Figures (\ref{fig:losssu2threemodels}), (\ref{fig:losssu4threemodels}) and (\ref{fig:su8multin1000nseg10}) show training and validation loss for the three models for the case of $\text{SU}(2)$, $\text{SU}(4)$ and $\text{SU}(8)$ for 1000 training examples, 10 segments, $h=0.1$ and 500 epochs. All models exhibit a noticeable flatlining of the MSE loss for after a relatively short number of epochs, indicative of the models saturating (reaching capacity for learning), a phenomenon accompanied by predictable overfitting beyond such saturation points. For small $h \approx 0.1$, the batch fidelity MSE is already at very low levels of the order $\sim 10^{-5}$. Again we see these persistently low MSEs as indicative of a highly determined model in which the task of learning $\Lambda_0$ (at least for smaller dimensional SU$(2^n)$) is overdetermined from the standpoint of large hidden layers (with 640 neuron units each), together with a prescriptive subRiemannian method.  From one perspective, these highly determined architectures such as SubRiemannian model have less applicability beyond the particular use-case of learning the subRiemannian geodesic approximations specified by the method in \cite{Swaddle_Noakes_Smallbone_Salter_Wang_2017}. A comparison of FC Greybox, which is a more generalisable architecture (not restricted to whitebox engineering of the subRiemannian algorithm) indicates relatively high performance measures of low MSE and high fidelity.


\begin{figure}[h]
\captionsetup[figure]{width=\textwidth}
\centering
\includegraphics[width=\linewidth]{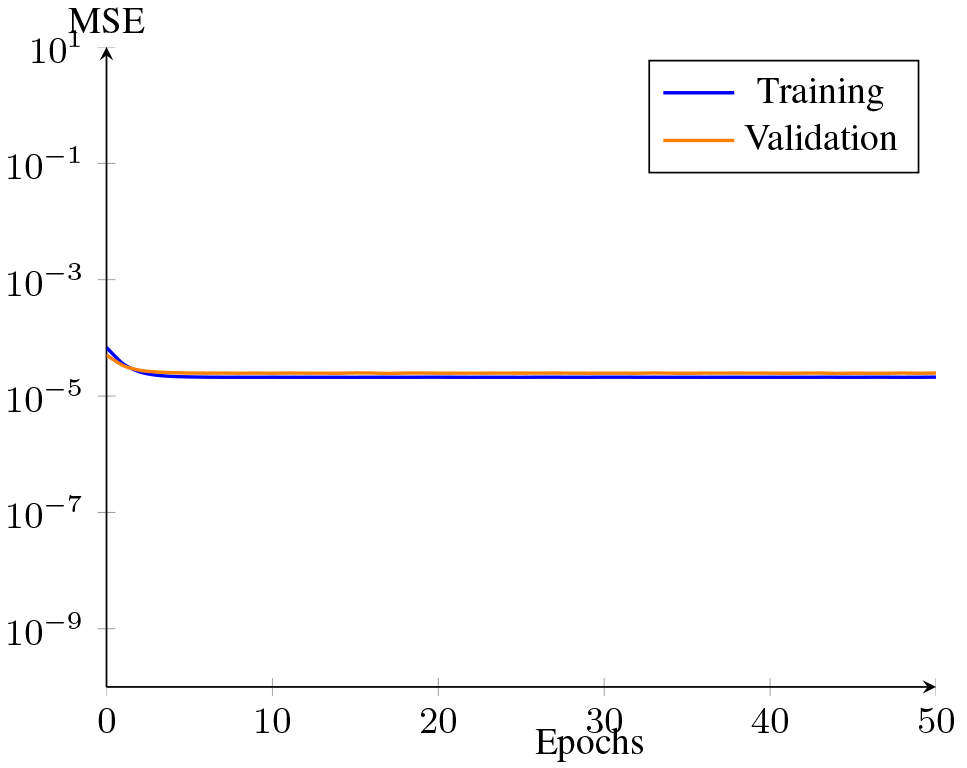}
\caption{Training and validation loss (MSE): GRU RNN Greybox. $G=\text{SU}(2),n_{\text{train}}=1000, n_{\text{seg}}=100,h=0.1$, epochs$=500$. This plot shows the MSE loss (for training and validation sets) for the GRU RNN Greybox model where the number of segments was increased from 10 to 100. As can be seen, the model saturates rapidly once segments are increased to 100 and exhibits no significant learning. Similar results were found for the SubRiemmanian model. This result suggests that simply changing the number of segments is insufficient for model improvement. One solution to this problem may be to introduce variable or adaptive hyperparameter tuning into the model such that the number of segments varies dynamically.}
  \label{fig:su2n1000subR}
\end{figure}

While the SubRiemannian model performed best in the case of $\Lambda_0 \in \frak{su}(2^n)$, as is evident from Tables (\ref{tab:mainresults}
 and \ref{tab:mainresultsdelta}), the GRU RNN Greybox model performed almost as well for SU$(2)$ and moderately outperformed the SubRiemannian model for SU(4) and SU(8) for most cases. The GRU RNN Greybox model was noticeably faster to train than the FC Greybox model by several hours and was slightly quicker to train than the SubRiemannian model but also flatlines (saturates) relatively early as evident in Figure (\ref{fig:su2n1000subR}). This is of note considering the fact that the GRU RNN Greybox model has more parameters than the SubRiemannian model (which ostensibly needs to only learn control amplitudes for $\Lambda_0$ generation) and is consistent with the demonstrable utility of GRU neural networks for certain quantum control problems \cite{Youssry_Chapman_Peruzzo_Ferrie_Tomamichel_2020}. One possible reason for differences between GRU RNN Greybox and SubRiemannian models may lie in the sensitivity of $\Lambda_0$: the SubRiemannian model's only variable degrees of freedom once initiated are in the relatively few weights $c^k$ learnt in order to synthesise $\Lambda_0$. As the dimension of $\text{SU}(2^n)$ grows, then the coefficients of $\Lambda_0$ become increasingly sensitive, that is, small variations in $c^k$ have considerable consequences for shaping the evolution in higher-dimension spaces, in a sense, $\Lambda_0$ bears the entire burden of training and so becomes hypersensitive and requires ever fine-grained tuning. This is in contrast to the GRU, for example, where the availability of more coefficients $c^k$ means each individual coefficient $c^k$ need not be as sensitive (can vary more) in order to learn the appropriate sequence. \\
 
\subsection{Segment and scale dependence}
\begin{figure}[h]
\captionsetup[figure]{width=\textwidth}
\centering
\includegraphics[width=\linewidth]{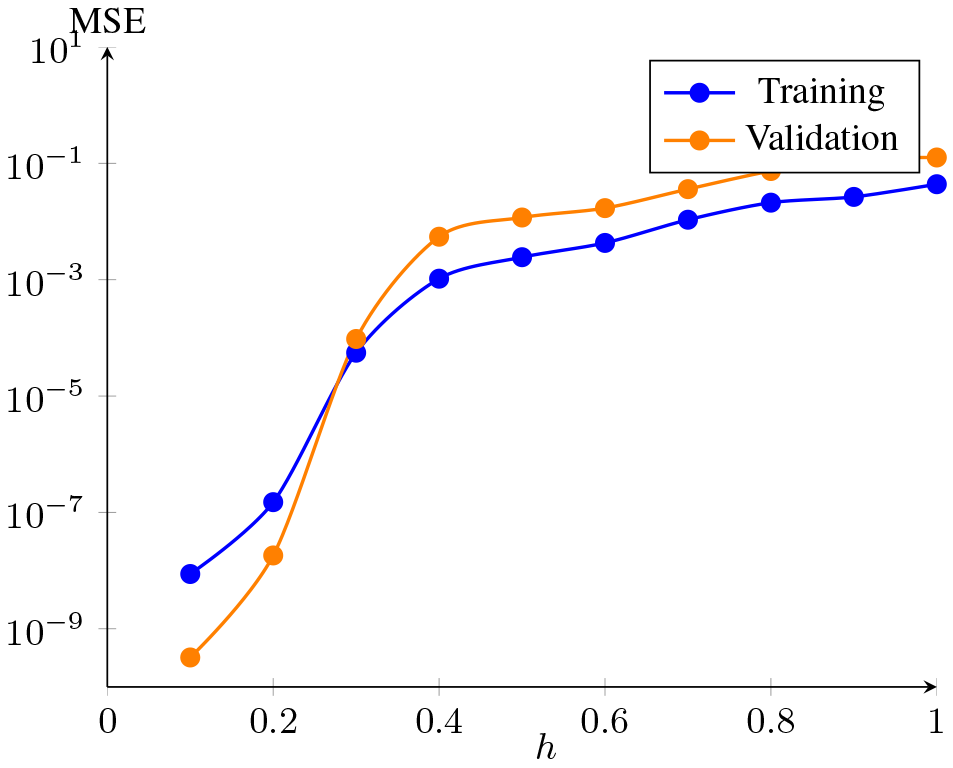}
\caption{Scale $h$ dependence (SubRiemannian model). $G=\text{SU}(2),n_{\text{train}}=1000, n_{\text{seg}}=10$, epochs$=500$. Plot demonstrates increase in batch fidelity MSE as scale $h$ ($\Delta t$) increases from 0.1 to 1, indicative of dependence of learning performance on time-interval over which subunitaries $Uj$ are evolved.}
  \label{fig:scaleplot}
\end{figure}

The experiments run across the various training sets indicated model dependence on the number of segments and scale $h$. As can be seen from Figure (\ref{fig:scaleplot}), we find that, not unexpectedly, the performance of models depends upon training data. In particular, model performance measures such as MSE and fidelity clearly depend upon time interval $h=\Delta t$: where $h$ is small, i.e. the closer the sequence of $(U_j)$ is to approximating the geodesic section, the lower the MSE and higher the fidelity. The effect on model performance is particularly evident in Figure (\ref{fig:lossscaledependent}) where increasing $h$ from 0.1 to 0.5 leads to a deterioration in loss of several orders in magnitude (particularly for $h>0.3$). As step size $h$ increases, the less approximating is the resultant curve to a geodesic. Furthermore, for larger step sizes, the conditions required for the assumption of time independence in unitary evolution (\ref{eqn:schrodind}) are less valid.

\subsection{Generalisation}
In order to test the generalisation of each model, a number of tests were run. In the first case, a set of random target unitaries $\tilde{U}_T$ from the relevant SU$(2^n)$ group of interest were generated. These target $\tilde{U}_T$ were then input into the SubRiemannian and GRU RNN Greybox models which output the estimated approximate geodesic sequences $(\hat{U}_j)$ to propagate from the identity to $\tilde{U}_T$. An estimated endpoint target estimate $\hat{U}_T$ for the approximate geodesic was generated by accumulating $(\hat{U}_j)$ i.e:
\begin{align}
    \hat{U}_n...\hat{U}_1 = \hat{U}_T.
\end{align}
This estimate $\hat{U}_T$ was then compared against $\tilde{U}_T$ to obtain a generalised gate fidelity $F(\tilde{U}_T,\hat{U}_T)$ for each test target unitary. Second, because fidelities of test unitary targets varied considerably, in order to test the extent to which higher fidelities may be related to similarity to the underlying training set of target unitaries $\{ U_T \}_{\text{train}}$ on which the models were trained, a second fidelity calculation was undertaken. The average gate fidelity of $\tilde{U}_T$ with $\{ U_T \}_{\text{train}}$ was calculated $\bar{F}(\tilde{U}_T,\{ U_T \}_{\text{train}})$. Correlations among the two fidelities were then assessed. 

In the third case, for SU(2) models trained on training data where $\Lambda_0 \in \Delta$, random test unitaries were replaced by $\tilde{U}_T$ comprising random-angle $\theta \in [-2\pi, \pi]$ $z$-rotations. The rationale was to test the extent to which a model based upon restricted control subalgebra training and architecture could replicate unitaries generated only from $\Delta$ with high fidelity for the single qubit case of $\text{SU}(2)$ where analytic solutions to the time optimal synthesis of subRiemanninan geodesics are known \cite{Boozer_2012}.  


\begin{figure}[h]
\captionsetup[figure]{width=\textwidth}
\centering
\includegraphics[width=\linewidth]{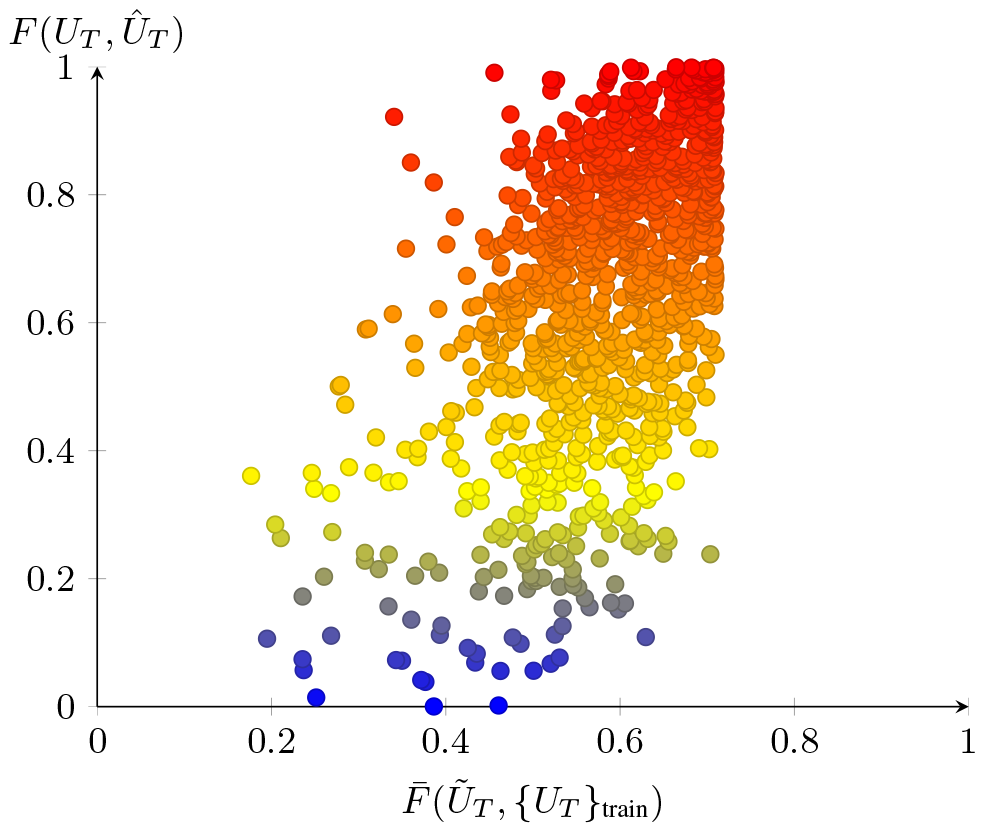}
\caption{Generalisation (SubRiemannian model). $G=\text{SU}(2),n_{\text{train}}=1000, n_{\text{seg}}=10$, epochs$=500$, $\Lambda_0 \in \frak{su}(2^n)$. Plot of generalised gate fidelity $F(\hat{U}_T,\tilde{U}_T)$  of randomly generated $\tilde{U}_T$ with the reconstructed estimate $\hat{U}_T$, versus $F(\hat{U}_T,\tilde{U}_T)$, average operator fidelity of randomly generated $U_T$ with training $\{U_T\}_{\text{train}}$ inputs to the model. The upward trend indicates an increase in operator fidelity as similarity (Pearson coefficient of $0.52$ to 95\% significance) of $U_T$ to training $\{U_T\}_{\text{train}}$ increases. Colour gradient indicates low fidelity (blue) to high fidelity (red).}
  \label{fig:su2fidvfidtrain}
\end{figure}


\begin{figure}[h]
\captionsetup[figure]{width=\textwidth}
\centering
\includegraphics[width=\linewidth]{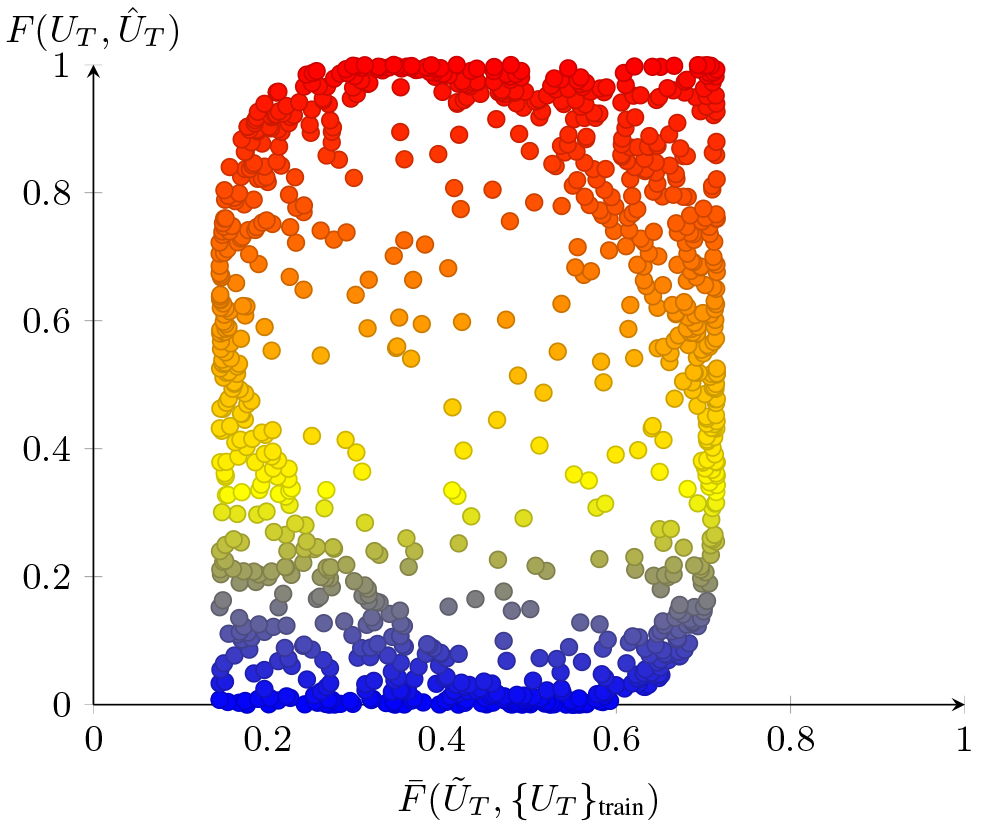}
\caption{Generalisation (SubRiemannian model). $G=\text{SU}(2),n_{\text{train}}=1000, n_{\text{seg}}=10$, epochs$=500$, $\Lambda_0 \in \Delta$. Plot of generalised gate fidelity $F(\hat{U}_T,\tilde{U}_T)$  of random-angle $\theta \in [-2\pi, \pi]$ $z$-rotations against generated $\tilde{U}_T$ with the reconstructed estimate $\hat{U}_T$, versus $F(\hat{U}_T,\tilde{U}_T)$, average operator fidelity of randomly generated $U_T$ with training $\{U_T\}_{\text{train}}$ inputs to the model. Here there is no statistically significant correlation between $U_T$ and training set $\{U_T\}_{\text{train}}$, though higher test fidelities are evident for $U_T$ bearing both high and low similarity to the training set (less dependence on similarity to training set for high fidelities).}
  \label{fig:su2fidvfidtraindelta}
\end{figure}


\twocolumngrid

Generalisation of both GRU RNN Greybox and SubRiemannian models trained on SU(2) was of mixed success. Figure (\ref{fig:su2fidvfidtrain}) plots $F(\tilde{U}_T,\hat{U}_T)$ against $\bar{F}(\tilde{U}_T,\{ U_T \}_{\text{train}})$ for the SubRiemannian model (comprising only $X,Y$ generators) trained on randomly generated unitaries in SU(2) where $\Lambda_0 \in \frak{su}(2^n)$ (colour gradient indicates low fidelity (blue) to high (red)). As can be seen, generalised gate fidelity varies considerably, with average generalised gate fidelity of 0.6474 with considerable uncertainty (standard deviation of 0.2288). In this case, there is a discernible relationship between fidelity and test unitary similarity to training data, evidenced by the upward trend of fidelities as similarity of $\tilde{U}_T$ to the training set $\{ U_T \}_{\text{train}}$ increases. The model was able to generate approximations to normal subRiemannian geodesics for certain random unitaries with a fidelity of over 0.99 by comparison with the intended target $\tilde{U}_T$. However, identifying structural characteristics among those estimates with higher fidelity remains an open problem. 

\begin{figure}[!t]
  \centering
  \includegraphics[width=\linewidth]{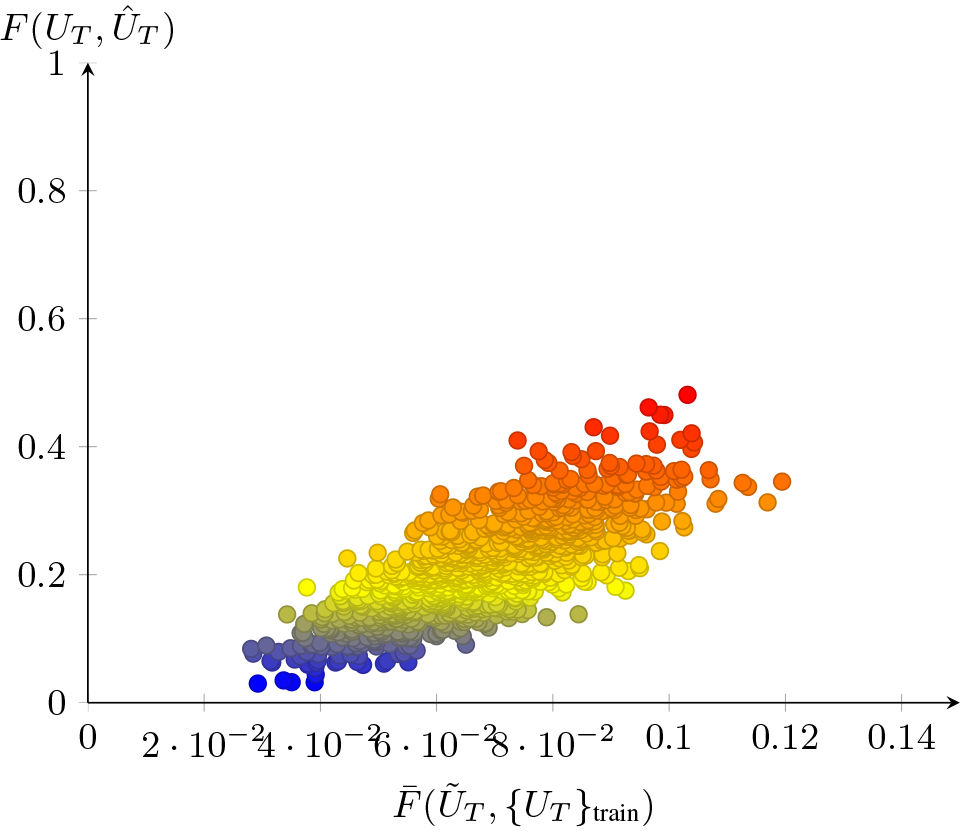}
\caption{Generalisation (SubRiemannian model).  $G=\text{SU}(8),n_{\text{train}}=1000, n_{\text{seg}}=10$, epochs$=500$, $\Lambda_0 \in \frak{su}(2^n)$. Plot of generalised gate fidelity $F(\hat{U}_T,\tilde{U}_T)$  versus $F(\hat{U}_T,\tilde{U}_T)$ (average operator fidelity against training set $\{U_T\}_{\text{train}}$). Generalisation was significantly worse for SU$(8)$, however correlation of generalised gate fidelity with similarity of $U_T$ to training sets is evident.}
  \label{fig:su8fidvfidtrain}
  \vspace*{\floatsep}
  \includegraphics[width=\linewidth]{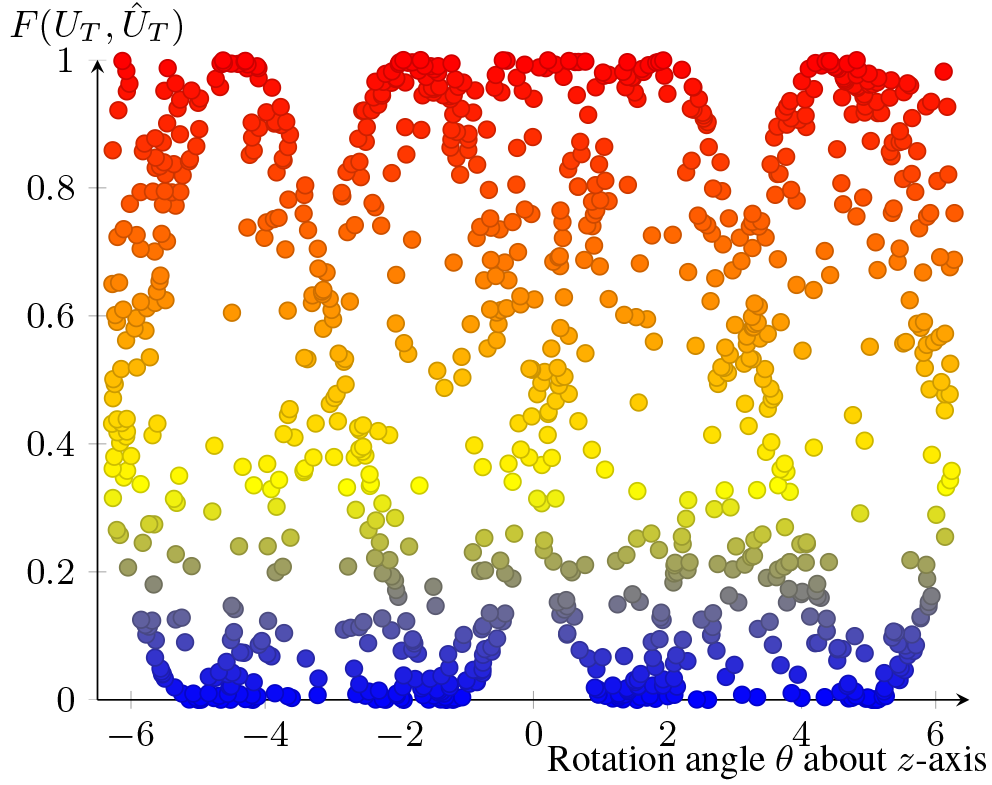}
\caption{Generalisation (SubRiemannian model). $G=\text{SU}(2),n_{\text{train}}=1000, n_{\text{seg}}=10$, epochs$=500$, $\Lambda_0 \in \Delta$. Plot of $F(\hat{U}_T,\tilde{U}_T)$  of random-angle $\theta \in [-2\pi, \pi]$ $z$-rotations $\theta$. As evident by the [red] high fidelities across the range $[-2\pi, \pi]$, the SubRiemannian model trained on data where $\Lambda_0 \in \Delta$ and $\Delta = \{X,Y \}$ in certain cases does generalise relatively well.}
  \label{fig:fidvanglezrotationsu2}
\end{figure}

By comparison, Figure (\ref{fig:su2fidvfidtraindelta}) plots the relationship between generalised gate fidelity and similarity to training set data for the SubRiemannian model trained on data generated where $\Lambda_0 \in \Delta$. In this case, the test unitaries $\tilde{U}_T$ were rotations by a random angle $\theta$ about the $z$-axis. No particular relationship between $\tilde{U}_T$ and the training set is apparent. Figure (\ref{fig:fidvanglezrotationsu2}) plots the same generalised gate fidelities in relation to $\theta$. Once again there is no immediately discernible pattern between the angle of the $z$-rotation and the fidelity of the estimate of $\tilde{U}_T$. We do see that high (above 0.99) fidelities are distributed across the range of $\theta$ and that there is some hollowing out of fidelities between extremes of 0 and 1.

The out of sample performance of both the SubRiemannian and GRU RNN Greybox models (in both cases limited to generators from $\Delta$) for random unitaries drawn from SU(4) and SU(8) was significantly worse than for SU(2). Average generalised gate fidelities were below 0.5 for each of the models tested. This is not unexpected given the heightened number of parameters that the models must deploy in order to learn underlying geodesic structure increases considerably as the Lie group dimension expands. A larger training set may have some benefits, however we note the saturation of the models suggests that at least for the models deployed in the experiments described above, expanding training sets is unlikely to systematically improve the generalisation of the models. Devising strategies to address both model saturation and ways in which expanded training data could be leveraged to improve model performance remains a topic for further research.






\section{\label{sect:conclusion}Conclusion and future directions}
This work  presents a comparative analysis of greybox models for learning and generating candidate quantum circuits from time optimally generated training data. The results from experiments above present a clear case for the benefits of greybox machine learning architecture for specific applications in quantum control. The increase in performance over blackbox models, as evidenced by training and validation average operator fidelities for synthesised quantum circuits of over 0.99 in each case, demonstrate that machine learning based methods of quantum circuit synthesis can benefit from customised architecture that engineers known or necessary information into learning protocols.
This is especially the case for quantum machine learning architectures: to the extent learning protocol resources need not be devoted to rediscovering known information or relationships, such protocols can more leverage the power of blackbox neural networks to those parts of problems which are unknown.

While the models outperformed current benchmarks on training and validation sets, they faced considerable challenges generalising well. Future work that may improve upon performance could include exploring hyperparameter learning, such as dynamically learning optimal numbers of segments, time-scale $h$ (including variable time-scale) or other metrics (such as Finslerian metrics) for use within the subRiemannian variational algorithm itself.
The cross-disciplinary intersection of geometry, machine learning and quantum information processing provides a rich seam of emergent research directions for which the application of both geometric quantum control and greybox machine learning architectures explored in this work  are potentially useful. 
It is important to note that the methods developed in this work , particularly the SubRiemannian model and GRU RNN Greybox were both specifically engineered for particular objectives. While the models developed in this work and experiments were tailored for the particular problem of learning subRiemannian normal geodesics for quantum circuit synthesis, the overall architectural framework in which geometric knowledge is encoded into machine learning protocols has potential for useful application in quantum and classical information processing tasks. 
Future work building upon the greybox machine learning results in this work could include an exploration of ways to combine the extensive utility of symmetric space formalism, methods of Cartan and other geometric techniques with machine learning.  
\\
\\
\textit{Acknowledgements}. The authors would like to acknowledge the assistance in particular of Akram Youssry and Christopher Sahadov Jackson in the preparation of this work. Elija Perrier is supported by an Australian Government Research Training Program Scholarship and Stipend from the Centre for Quantum Software and Information at UTS. Funding for this work was provided by the Australian Government via the AUSMURI
grant AUSMURI000002.


\newpage
\twocolumngrid
 \bibliography{apssamp}

\appendix

\section{Algorithmic architectures}
\label{A0:Algo}
The section below sets-out pseudocode for the machine learning models utilised in the experiments above.
\subsection{Fully-connected Greybox model}
\label{A0:FCG}
Pseudocode for the Fully-connected Greybox model is set-out below. Note that TensorFlow inputs required $(U_j)$ to be separated into real $\text{Re}(U_J)$ and imaginary $\text{Im}(U_J)$ parts and then recombined for input into fidelity calculations. Note the cost function $C(F,1)$ below is implicitly a function of $c^k$ (the sequence of which is $(c_j)$) which are the variable weights in the model. Here $\gamma$ is the learning rate for the gradient update and $\theta$ the trainable weights of the model.
\begin{algorithm}[H]
\SetAlgoLined
 Inputs: $U_T$, $\text{Re}(U_J)$, $\text{Im}(U_J)$,$\Delta$,$h$\\
 Labels: $v = (1...1)$, $\dim v = |(U_j)|$ \\
 FC Dense Network: $U_T \to \tanh(U_T;\theta) = (\hat{c}_j)$\\
 Hamiltonian estimation: $(\hat{c}_j), \Delta \to (\hat{H}_j) = (\sum_k\hat{c}^k \tau_k)$ where $\tau_k\in \Delta$\\
 Quantum Evolution: $(\hat{H}_j) \to (\hat{U}_j) = (\exp(-h \hat{H}_j))$\\
 Fidelity: $\text{Re}(U_J),\text{Im}(U_J),(\hat{U}_j) \to F(\hat{U}_j, U_j)$\\
 MSE: $\min C(F,1) = \frac{1}{n}\sum_j^n (1 - F(\hat{U}_j,U_j))^2$\\
 Update: $\theta \to \theta - \gamma\nabla_{\theta} C(F,1)$
 \caption{Fully-connected Greybox model}
\end{algorithm}

\subsection{GRU RNN Greybox model, parameters $\theta = (w,b)$}
\label{A0:GRULSTM}
Pseudocode for the GRU RNN Greybox model is set-out below. Note that TensorFlow inputs required $(U_j)$ to be separated into real $\text{Re}(U_J)$ and imaginary $\text{Im}(U_J)$ parts and then recombined for input into fidelity calculations. Note the cost function $C(F,1)$ below is implicitly a function of $c^k$ (the sequence of which is $(c_j)$) which are the variable weights in the model. Here $\gamma$ is the learning rate for the gradient update and $\theta$ the trainable weights of the model.

\begin{algorithm}[H]
\SetAlgoLined
 Inputs: $U_T$, $\text{Re}(U_J)$, $\text{Im}(U_J)$,$\Delta$, $h$\\
 Labels: $v = (1...1)$, $\dim v = |(U_j)|$ \\
 GRU RNN: $U_T \to \tanh(U_T;\theta) = (\hat{c}_j)$\\
 Hamiltonian estimation: $(\hat{c}_j), \Delta \to (\hat{H}_j) = (\sum_k\hat{c}^k \tau_k)$ where $\tau_k\in \Delta$\\
 Quantum Evolution: $(\hat{H}_j),h \to (\hat{U}_j) = (\exp(-h \hat{H}_j))$\\
 Fidelity: $\text{Re}(U_J),\text{Im}(U_J),(\hat{U}_j) \to F(\hat{U}_j, U_j)$\\
 MSE: $\min C(F,1) = \frac{1}{n}\sum_j^n (1 - F(\hat{U}_j,U_j))^2$\\
 Update: $\theta \to \theta - \gamma\nabla_{\theta} C(F,1)$
 \caption{GRU RNN Greybox model}
\end{algorithm}

\subsection{SubRiemannian model}
\label{A0:SUB}
Pseudocode for the SubRiemannian model is set-out below. Note that TensorFlow inputs required $(U_j)$ to be separated into real $\text{Re}(U_J)$ and imaginary $\text{Im}(U_J)$ parts and then recombined for input into fidelity calculations. Note the cost function $C(F,1)$ below is implicitly a function of $c^k$ (the sequence of which is $(c_j)$) which are the variable weights in the model. Here $\gamma$ is the learning rate for the gradient update and $\theta$ the trainable weights of the model.
\begin{algorithm}[H]
\SetAlgoLined
 Inputs: $U_T$, $\text{Re}(U_j)$, $\text{Im}(U_j)$,$A = \Delta$ or $\frak{su}(2^n)$, $U_0=I$, $h$, $n_{\text{seg}}$ \\
 Labels: $v = (1...1)$, $\dim v = |(U_j)|$ \\
 FC Dense Network: $U_T \to \tanh(U_T; \theta) = c_{\Lambda_0}^{+}, \tanh(U_T; \theta) =c_{\Lambda_0}^{-}$\\
 $\Lambda_0$ estimation:\\
 \qquad $c_{\Lambda_0}^{+}, A \to \Lambda_0^{+} = \sum_k c^{k+} \tau_k$, $\tau_k \in A$;\\
 \qquad $c_{\Lambda_0}^{-}, A \to \Lambda_0^{-} = \sum_k c^{k-} \tau_k$, $\tau_k \in A$\\
 $\Lambda_0$ layer: $\Lambda_0^{+},\Lambda_0^{-} \to \Lambda_0$ \\
 subRiemannian layer: $\Lambda_0 \to (U_j)$. Set $Y = U_0$.\\ 
 For $j$ in $n_{\text{seg}}$:\\
 \qquad $\Lambda_0 \to Y \Lambda_0 Y^\dagger = X$\\
 \qquad $X \to \hat{H}_j=\projdelta(X)$, $c_j$\\
 \qquad $\hat{H}_j \to \hat{U}_{j+1}=\exp(-h H_j)$\\
 \qquad $Y = \hat{U}_{j+1}$\\
 \qquad return $(\hat{U}_j)$\\
 Fidelity: $(\hat{U}_j),(\text{Re}(U_j)),(\text{Im}(U_J)) \to F(\hat{U}_j, U_j)$\\
 MSE: $\min C(F,1) = \frac{1}{n}\sum_j^n (1 - F(\hat{U}_j,U_j))^2$\\
 Update: $c^k \to c^k - \gamma\nabla_{c^k} C(F,1)$
 \caption{SubRiemannian model}
\end{algorithm}

\subsection{Simulation Design}
\label{A0:Simulation}
Simulation of training datasets for use in the machine learning models was undertaken in Python. The simulation was adapted from Mathematica code accompanying \cite{Swaddle_Noakes_Smallbone_Salter_Wang_2017} with a number of adaptations. The code is constructed as a class with the following hyperparameters: (i) $n=\dim(\text{SU}(2^n))$ for selecting the Lie group of interest $\text{SU}(2^n)$; (ii) $n_{\text{seg}}$ the number of segments (indexed by $j$) in the global decomposition into subunitaries $(U_j)$; (iii) $n_{\text{train}}$, the number of training examples; (iv) a parameter for whether $\Lambda_0 \in \frak{su}(2^n)$ or $\Delta$; (v) a set of parameters for selecting (for $\text{SU}(2)$) which Pauli operators constituted $\Delta$; (vi) a parameter for selecting whether unitaries were to be generated in accordance with the example formulation in \cite{Boozer_2012}, (vi) parameter for selecting $h$ (which defaults to $1/n_{\text{seg}}$ in the event of a null entry. Upon selection of parameters, the class generates an extensive selection of training data in various forms (see the relevant code repository for code with commentary), including complex and realised iterations of $U_T, (U_j), c_j, \Delta$ and other key inputs into the models. Training data was generated using a combination of standard Python numerical and scientific packages together with quantum simulation software QuTip \cite{Johansson_Nation_Nori_2013}. 


\section{Differential geometry and Lie groups}
\label{A1:Differential}
\subsection{Generating subalgebras for geodesics}
\label{A11:onetwobodyoperators}
\subsubsection{Product Operator Basis}

 Our experimental results and methods focus on synthesising quantum circuits for multi-qubit systems where unitary operators are drawn from $\text{SU}(2^n)$. For such multi-qubit (qudit) systems, unitary operators $U$ belong to Lie groups $G = \sutwon$ which describe the evolution of $n$ interacting spin$-1/2$ particles. These groups are equipped with a corresponding Lie algebra of dimension $(2^n)^2-1=4^n-1$ and denoted $\liesutwon$, represented via traceless $n \times n$ skew-Hermitian ($A = -A^*$) matrices. Solving time-optimal problems in such contexts often relies upon appropriate selection of a subset of generators from the Lie algebra as the control subalgebra from which to synthesise a quantum circuit. This is especially the case when selecting a control algebra that renders targets $U_T$ reachable in a way that approximates geodesic curves on the relevant manifold as the choice of one set of generators over another can affect evolution time (and whether generated geodesics are indeed minimal, in cases where multiple geodesics are available such as via great circles on a 2-sphere).    
 Of importance in selecting control subalgebras for time-optimal synthesis of geodesics in multi-qubit systems \cite{Nielsen_Dowling_Gu_Doherty_2006, Gu_Doherty_Nielsen_2008, WangLloyd2015, Khaneja_Glaser_2001, DAL2008}
 is the so-called \textit{product operator basis} i.e. a basis for the Lie algebra of generalised Pauli matrices, being tensor (Kronecker) products of elementary Pauli operators. The basis is formed by Pauli spin matrices $\{I_x, I_y, I_z\} = 1/2\{\sigma_x, \sigma_y, \sigma_z\}$ i.e. the sets of generators of rotation in two-dimensional Hilbert space (and Lie algebra basis), with usual commutation relations. A basis for $\sutwon$ comprises of many-body tensor products of these Pauli operators, i.e. for an $n$-dimensional operator, there are between $1$ and $n$ Pauli operators tensor products with identities for various indices. An orthogonal basis $\{iB\}$ (frame) for $\liesun$ is then given \cite{Khaneja_Glaser_2001} in closed-form via:
\begin{align*}
    B_s = 2^{q-1} \Pi_{k=1}^n (I_{k\alpha})^{a_{ks}}
\end{align*}
where $\alpha =x,y,z$ and
\begin{align*}
    I_{k\alpha} = 1 \otimes ... \otimes I_\alpha \otimes 1
\end{align*}
where $I_\alpha$ appears only at the $k$th position with the identity appearing everywhere else.
The parameter $q$ tells us how many Pauli operators are tensor producted together e.g. $q=1$ means the basis element only has one Pauli and the rest identities; $q=2$ means we are dealing with a basis formed by tensor products of two Pauli operators and identities etc.

\subsubsection{One- and two-body operators}
Geometric control techniques for multi-qubit systems often focus on selecting one- and two-body Pauli product operator frames (bases) for relevant control subalgebras \cite{Dowling_Nielsen_2008, SwadThesis}. If the control subalgebra contains only one- and two-body elements of the Lie algebra $\frak{g}$, then curves generated in the corresponding Lie group $G$ are more likely (with a number of important caveats) to be approximations to (and in the limit, as the number of gates $n \to \infty$, representations of) geodesic curves and thus time-optimal synthesis of target unitary propagators. This approach can be seen across a number of key results in the literature \cite{Khaneja_Glaser_2001, Dowling_Nielsen_2008, Gu_Doherty_Nielsen_2008, WangLloyd2015} and forms the basis for the relevant distribution used in subRiemannian variational methods in \cite{Swaddle_Noakes_Smallbone_Salter_Wang_2017,SwadThesis} which the protocols developed in this workexpand upon. The preference for one- and two-body Pauli operator frames arises in different contexts.  

For example, it is demonstrated in \cite{Khaneja_Glaser_2001} in the case where $G=\text{SU}(4)$ and $K = \text{SU}(2) \otimes \text{SU}(2)$ that by finding an appropriate Cartan decomposition $G=KAK$ (with associated Lie algebra decomposition $\frak{g} = \frak{k} \oplus \frak{p}$) and maximally abelian subalgebra
\begin{align*}
    \frak{h} = i\spn{I_xS_x, I_yS_y, I_zS_z} \subset \frak{p}
\end{align*}
(where $I_\alpha$ represent one-body terms and $S_\beta$ two-body terms), we can write $\exp(-i\frak{h}) = A$ in the $KAK$ decomposition as the exponential of a linear combination of the generators in $\frak{h}$, namely:
\begin{align*}
    U_F = K_1 \exp(-i(\alpha_1I_xS_x + \alpha_2I_yS_y + \alpha_3I_zS_z  )) K_2
\end{align*}
where $K_1,K_2 \in K = \sutwo \otimes \sutwo$. In this case, any Hamiltonian from from $\frak{h} \subset \frak{p}$ can be generated using the controls in $K$ (essentially by showing they can generate the two-body terms via action of the single-body operators $I$ on $S$) and is time optimal. Because synthesis depends on the evolution of the drift Hamiltonian according to generators in $\frak{p}$ (as acted on via the adjoint action of $K$) and because this depends on the coefficients of the generators $\alpha_i$, then the minimal time is given by the coefficients of the generators in $\frak{k}$ used to steer $H_d$:
\begin{align*}
    T&=\min_{\alpha_i} \sum_{i=1}^3 |\alpha_i|
\end{align*}
hence the optimisation problem becomes relatively straight-forward. One rationale for preferring one- and two-body generators \cite{Khaneja_Glaser_2001} is that higher-order (i.e. more than one-body) generators are shown to have coefficients which include a scalar coupling strength $J$ between the relevant spins such that each $H_j$ has a coefficient $2\pi$ and the two-body ($I_iS_i$) term has coefficient $2\pi J$. Thus a time-optimal problem becomes a simpler optimisation problem of finding the minimal sum $\sum_i \alpha_i$ satisfying:
\begin{align*}
    U_F = Q_1 \exp(-i2\pi J(\alpha_1I_xS_x + \alpha_2I_yS_y + \alpha_3I_zS_z)) Q_2
\end{align*}
where $Q_1,Q_2 \in K$. The proof essentially relies on the fact that because synthesis of $Q_1,Q_2$ takes negligible time, then synthesis time is determined by the time to synthesise $A$ in the $KAK$ which is determined by the parameters $\alpha_i$, hence minimal time amounts to minimising the sum of $\alpha_i$. Synthesis time is thus minimal to the extent that the `fewest-body' Pauli generators are utilised in the control subalgebra. Thus, ideally, to generate minimal (and thus time optimal) paths in $G$ to reach arbitrary target unitaries $U_T$, one should ideally choose the control subalgebra with as few many-body terms as necessary in order to render $U_T$ reachable in a control sense. 

\subsubsection{Nielsen's approach}
\label{B1:Nielsen}
Nielsen et al. also focus on adopting one- and two-body terms in their metric-based approach to characterising and generating time-optimal quantum circuits. For example, the preference for one- and two-body generators is justified in  \cite{Nielsencomplex} imposing  a \textit{Hamming weight} term $\text{wt}(\sigma)$ applied to the Pauli generators $\sigma$ together with a penalty function $p(\cdot)$ that penalises the control functional whenever Pauli terms of high Hamming weight are part of the control Hamiltonian. The idea is that Pauli $n$-tuples (tensor products) of anything more than one- or two -body Hamiltonians will be penalised via a higher Hamming weight as they will have many more non-identity elements, whereas one- and two-body operators have lower Hamming weight). Nielsen et al. demonstrate that selection of one- and two-body generators is optimal for calculating a lower bound on the complexity measure $m_\mathcal{G}(U)$ using Finsler metrics i.e:
\begin{equation}
    d_F(I,U) \leq m_\mathcal{G}(U)
\end{equation}
 where $\mathcal{G}$ is a universal gate set in $\sutwon$. 

The significance of restricting control subalgebras together with bespoke metrics when utilising geometric optimisation techniques is evident in later work \cite{Nielsen_Dowling_Gu_Doherty_2006}. For quantum control optimisation architectures, this demonstrates the utility of Finsler metrics as a more general norm-based measure of distance (and thus optimality) together with a justification of the selection of one- and two-body generators due on the basis of Hamming weights. The use of the `penalty metric' approach is explored in further work \cite{Gu_Doherty_Nielsen_2008, WangLloyd2015} however, as noted in \cite{SwadThesis}, such approaches can be convoluted without providing guarantees that optimal generators will be selected.

In \cite{Nielsen_Dowling_Gu_Doherty_2006}, Nielsen et al. expand certain elements of the initial program combining techniques from differential geometry and variational methods to quantum circuit synthesis and quantum control. This second paper considered the difficulty of implementing a unitary operation $U$ generated by a time dependent Hamiltonian evolving to the desired $U_T$. They show that the problem of finding minimal circuits is equivalent to analogous problems in geometric control theory i.e. this paper has more of a focus on quantum control utilising geometric means. They select a cost function on $H(t)$ such that finding optimal control functions for synthesis of $U_T$ (evolving according to the Schrodinger equation) involves finding minimal geodesics on a Riemannian manifold.

In this case, $H(t)$ is written in terms of a \textit{Pauli operator expansion}:
\begin{equation}
    H = \sum_\sigma^{'} h_\sigma \sigma + \sum_\sigma^{''} h_\sigma \sigma
\end{equation}
where the first summation is over one- and two-body terms, the second over all other tensor products. A cost function is constructed with a penalty term $p^2$ imposed that penalises the higher-order terms
\begin{equation}
    F(H) = \sqrt{\sum_\sigma^{'} h_\sigma \sigma + p^2\sum_\sigma^{''} h_\sigma \sigma}
\end{equation}
with the total cost to be minimised given by
\begin{equation}
    d([U]) \equiv \int_0^T dt F[H(t)]
\end{equation}
Due to parametrisation invariance, $F$ (a Finsler metric) can be rescaled such that $T = d([U])$. 
The overall effect is to demonstrate that using $O(n^2 d(I,U)^3)$ one- and two-qubit gates, it is possible to synthesise a unitary $U_A$ satisfying $||U_T - U_A|| \leq c$, where $c$ is a constant and $U_T$ is the target unitary gate. Moreover, the work demonstrates the optimality of unitary synthesis via following minimal geodesics in the Lie group manifold generated by one- and two-body generators (as we focus on below). Nielsen notes the number of one- and two-qubit terms (i.e. $\dim \Delta$) for the relevant Lie algebra is given by
\begin{equation}
    \dim \Delta=9n(n-1)/2 + 3n
\end{equation}
a relatively trivial but important feature of the machine learning code in model architectures explored below.

 Later work \cite{Dowling_Nielsen_2008} of Nielsen and Dowling provides a more directly applicable example of how to develop analytic solutions to geodesic synthesis of unitary operations. As with the discussion above, it is worth exploring the key results from this work in order to understand characteristics of relevance to any attempt to utilise geometric methods for synthesis of unitary propagators for multi-qubit systems. In the paper, they develop a method of deforming (homotopically) simple and well-understood geodesics to geodesics of metrics of interest. Intuitively, the idea is to start with a known geodesic curve between $I$ and $U_T$ and, subject to certain constraints, `bend' it homotopically (that is, via mappings which preserve topological properties) into a minimal-length curve. However, as demonstrated in \cite{Dowling_Nielsen_2008}, a similar preference for one- and two-body terms is manifest in the applicable lifted Hamilton-Jacobi equation (this paper is also important for anyone interested in geometric quantum control given its discussion of significant (and potentially intractable) complexity constraints presented by the quantum extension of the Rabarov-Rudich theorem and also extend geometric quantum computing to include ancilla qubits.

In subsequent work utilising Nielsen et al.'s approach \cite{WangLloyd2015}, the application of penalty metrics is extended in order to show its utility in synthesising time-optimal geodesics. In that paper, it is shown that a bound on the norm of the Hamiltonian $H(t)$ is equivalent to a bound on the speed of evolution, that is, such a bound implies that minimal-time paths are minimal distance in which the norm function is used as the distance measure. Given $||H(t)||=E$, they demonstrate that for any curve connecting $I$ and $U_T$, the length of time-optimal curves is given by:
\begin{align}
    L = \int_0^T||H(t)||dt = \int_0^T Edt = ET
\end{align}
where minimising evolution time $T$ thereby minimises distance $L$. Hamiltonians of interest are confined to a control subalgebra $\mathcal{A}$ that disjunctively partitions the Lie algebra $\mathcal{M}$ (i.e. equivalent to generators being drawn from $\frak{k}$ or $\frak{p}$ above) and cases where $||H(t)|| \leq E$ (where the Hamiltonian can rescaled i.e. reparametrised so that the norm equals $E$ at all points on the path which in effect keeps the path identical but time shorter). They introduce a slight modification, that $\trace(H^2(t)) = E^2$ in order to introduce the \textit{quantum brachistorone problem} (see also \cite{Carlini_Hosoya_Koike_Okudaira_2007}), a quantum analogue of the brachistorone (meaning `shortest time') problem from classical variational mechanics \cite{Goldstein_2002}.    
Their method in essence adopts the penalty-metric approach of \cite{Dowling_Nielsen_2008} such that in the limit, the lowest-energy solution tends towards minimal time by reason of the increased cost associated with higher-order (more than one- and two-body) generalised Pauli generators. 

The approach in \cite{WangLloyd2015} is precisely to use the penalty metric approach of Nielsen et al. to generate a subRiemannian geodesic equation in order to confine the generators of the curve on the manifold to the control subalgebra $\mathcal{A}$. This is achieved by adopting the norm-based cost function (pseudometric) where higher-order generator terms are weighted with penalty $q$, so that minimisation will by extension favour those generators (i.e. favour generators in $\frak{k}$ not $\frak{p}$). By doing so, a sufficiently proximal initial seed for the ``shooting method'' (see \cite{WangLloyd2015, Press_Teukolsky_Vetterling_Flannery_2007}) is generated. This method is a generic numerical technique for solving differential equations with two-point boundary problems (where our two points are $I$ and $U_T$ on $G$) and thus generating approximate geodesics.

 Another motivation of restricting control subalgebras as in \cite{SwadThesis, Swaddle_Noakes_Smallbone_Salter_Wang_2017} and our experiments below) to one- and two-body terms is to be found via the  geodesic approximations via the decomposition of the Lie algebra into projective subspace operators \cite{Dowling_Nielsen_2008, Brandt_2010a, Brandt_2010b}. In \cite{WangLloyd2015}, this was achieved via setting $\mathcal{P}(H) = H_P$ for one- and two-body Pauli terms and $\mathcal{Q}(H) = H_Q$ for three- or more-body Pauli terms such that
\begin{equation}
    \liesutwon = \mathcal{P} + \mathcal{Q} \qquad H = H_P + H_Q
\end{equation}
The idea is that higher-order (three- or more-body) terms in $\{ H_Q \}$ carry a penalty parameter (weight) which is designed, when curve length is obtained via minimising the action, to penalise higher-order terms in a way that the functional (solution) to the variational problem is more likely to contain only one- and two-body terms. Thus instead of restricting the sub-algebra of controls $\frak{k}$ to only one- and two-body terms (such as is undertaken in Swaddle), they instead (as per Nielsen's original paper) begin with full access to the entire $\liesutwon$ Lie algebra (i.e. fully controllable) and then proceed to impose constraints in order to refine this down to geodesics comprising only (or mostly) one- and two-body terms. The distinction with the subRiemannian approach adopted in \cite{Swaddle_Noakes_Smallbone_Salter_Wang_2017} is that in the latter case, generators for $U$ are by design constrained to be drawn from $\mathcal{P}=\Delta$ via the projection function (\ref{eqn:projection}), circumventing imposition of Finslerian penalty metrics.

\section{Comparing geodesic approximations}
\label{A:boozerswaddlecomparison}
Generation of geodesics in a QML context relies upon the availability of ways to compare whether outputs of machine learning models do in fact closely approximate geodesic curves. Thus the availability of reliable analytic and numerical methods for the generation of geodesics for use as training, testing and validation datasets is important. 
In our work, we sought to adapt the novel algorithmic approach to geodesic synthesis from \cite{Swaddle_Noakes_Smallbone_Salter_Wang_2017} to include performance metrics of relevance to quantum information processing, such as fidelity measures. By comparison, \cite{Boozer_2012} sets out an algorithm for determining time-optimal sub-Riemannian geodesics in SU($2$) which can be used to benchmark the performance of different machine learning approaches to synthesising approximate geodesics. While the derivation of time optimal parameters in \cite{Boozer_2012} relies upon complicated sequence of coordinate transformations which is not easily scalable, it does provide a useful basis for comparison with the methods in \cite{Swaddle_Noakes_Smallbone_Salter_Wang_2017}. 

In \cite{Boozer_2012}, it is shown that time-optimal paths, where target unitaries constitute rotations by angle $\theta$ about the $z$-axis with generators being Pauli $X$ and $Y$ operators, can be synthesised in time-optimal fashion by following `circular' or holonomic paths along which they are parallel-transported. On the Bloch sphere, this is represented as `circular' paths emanating from the north pole whose diameter increases with increasing $\theta \in [0,2\pi]$. Intuitively, the greater the angle of rotation, the greater the diameter of the holonomic path. 

To this end, one of the objectives of our experiments was to ascertain the reliability of a few different methods of generating geodesics using methods drawn from geometric control sources. In order to do so, we compared this variational geodesic generation \cite{Sachkov_2009} approach to a known method for analytically determining subRiemannian geodesics in SU($2$) in \cite{Boozer_2012}. By demonstrating the existence of a homeomorphism between the two methods one can be confident that the variational method appropriately approximates geodesics.

The challenge posed in comparing geodesic methods lies in the differing assumptions of each method: \cite{Swaddle_Noakes_Smallbone_Salter_Wang_2017} constrains the norms $||\projdelta(u_0)||=||u_0||$ as a means of more efficiently generating subRiemannian geodesic approximations \cite{SwadThesis}, which is in effect the time scale (or energy scale) of their method. Conversely, \cite{Boozer_2012}, works at different scales. In practice this means the generators for unitary evolution via each method differ by a scaling related to the norm of the generators. Such different parameterisations can be understood as follows:
\begin{align*}
    &\text{Swaddle parametrisation} & \text{Boozer parametrisation}\\
    &||H_j^{(S)}||=\Omega_j &  ||H_j^{(B)}||=1\\
    &dt_j^{(S)} = h = 1/N & dt_j^{(B)} = \Omega_j h/1\\
    &t_j^{(S)} = \sum_{k=1}^j h = jh & t_j^{(B)}  = \sum_{k=1}^j\Omega_j h/1
\end{align*}
For some desired tolerance (difference) $\epsilon$, the two approximations at are identical if the cumulative norms $D(H^{(S)},H^{(B)})$ of the sum of their $j$th Hamiltonians satisfy:
\begin{align}
    D(H^{(S)},H^{(B)}) & = \sum_j \left| \left| \frac{H_j^{(S)}}{\Omega_j} - H_j^{(B)} \right| \right| < \epsilon. \label{eqn:swadboozmetric}
\end{align}
That is, we want to minimise the distance between each Hamiltonian segment. The result in \cite{Boozer_2012} is a relatively simple control problem where the control subalgebra consists of Pauli $\sigma_x, \sigma_y$ generators with the target a rotation about the $z$-axis by angle $\eta$, $U_T = \exp(-i\eta \sigma_z/2)$. To validate that variational subRiemannian method can reproduce the time-optimal paths from \cite{Boozer_2012}, a transformation between the two that enables comparison of Hamiltonians at time $t_j$ respectively in each formulation must be found. Pseudocode for such a transformation (in effect, a rescaling) of Hamiltonians generated using the method in \cite{Swaddle_Noakes_Smallbone_Salter_Wang_2017} by comparison with those using the method in \cite{Boozer_2012} is set-out below (where $(S)$ indicates Hamiltonians using the method in \cite{Swaddle_Noakes_Smallbone_Salter_Wang_2017} and $(B)$ the method in \cite{Boozer_2012}).

\begin{algorithm}[H]
\SetAlgoLined
 Generate $H_j^{(S)}$\\
 Calculate $||H_j^{(S)}|| = \Omega_j$\\
 Calculate $t_j^{(B)}  = \sum_{k=1}^j\Omega_j h$\\
 $H_0^{(B)} = \frac{1}{\Omega_j}H_1^{(B)}$\\
 Calculate $H_j^{(B)} = e^{-i\omega t_j^{(B)} \frac{\sigma_z}{2}} H_0^{(B)} e^{i\omega t_j^{(B)} \frac{\sigma_z}{2}}$
 \caption{Comparison of subRiemannian and analytic geodesic circuits in $\text{SU}(2)$}
\end{algorithm}
Here, conjugation by $\exp(-i\omega t_j^{(B)} \frac{\sigma_z}{2})$ represents the Euler decomposition of the evolution in \cite{Boozer_2012} as if one had direct access to the generator $\sigma_z$. Alternatively, one can also compare unitaries at equivalent times via operator fidelity $F(U_j^{(B)},U_j^{(S)})$ where:
\begin{align}
    U_j^{(B)}&=\exp(-i H_j^{(B)} dt_j^{(B)})
\end{align}
where we again use the assumption:
\begin{align}
    U_F^{(B)}&\approx e^{-iH_N^{(B)} dt_N^{(B)}}... e^{-iH_1^{(B)} dt_1^{(B)}}.
\end{align}
Numerical results comparing both Hamiltonian average distance (\ref{eqn:swadboozmetric}) and fidelities for ten $U_j$ instances across $N$ segments are set-out below.

\begin{center}
 \begin{table}[t]
 \begin{tabular}{|c|| c| c |} 
 \hline
 $j$ & $D(H^{(S)},H^{(B)})$ & $F(U_j^{(B)},U_j^{(S)})$ \\
 \hline
 \hline
 1 & 0.0922 & 0.9934 \\
 2 & 0.1986 & 0.9935 \\
 3 & 0.3105 & 0.9935 \\
 4 & 0.4169 & 0.9936 \\
 5 & 0.5154 & 0.9936 \\
 6 & 0.6046 & 0.9936 \\
 7 & 0.6836 & 0.9937 \\
 8 & 0.7518 & 0.9937 \\
 9 & 0.7871 & 0.9938 \\
 10 & 0.8345 & 0.9939\\
 \hline
 \end{tabular}
 \caption{Hamiltonian distance and unitary fidelity between Swaddle and Boozer geodesic approximations. Fidelity}
\label{table:1}
\end{table}
 \end{center}
Fidelity results indicate little difference between $U_j^{(S)}$ and $U_j^{(B)}$, while Hamiltonian distance increases with $j$. Overall, the results provide some measure of confidence, though not analytic certainty, that the variational subRiemannian means of geodesic approximation in \cite{Swaddle_Noakes_Smallbone_Salter_Wang_2017} are useful candidates for training data.

\newpage
\section{Neural network and GRU architectures}
\label{A2:GRUNN}
\subsection{Feed-forward neural networks}
Feed-forward fully-connected neural networks, such as the ones deployed in the models above, can be understood in terms of functional composition. The objective of deep feed-forward networks is to define an input-output function $z = f(a,w,b)$ where $a^l$ are inputs to the layer $l$ (setting the initial input $a^0=x$), $w^l$ is a tensor of parameters for layer $l$ to be learnt by the model and $b^l$ is a bias tensor applied to $a^l$ \cite{Nielsen_2015, Goodfellow_Bengio_Courville_2016}. 

In its simplest incarnation, the feed-foward stack takes as input a flattened realised $a_0 = U_T$ (where $k$ runs over the dimension of the vector). A layer of a simple neural network consists of units or neurons activation functions $\sigma$ (in our case, the ReLU or tanh activation function) applied to the $z$ such that we have $a^l=\sigma(z^l)$, vector and bias $b$:
\begin{align}
    a^l=\sigma(z^l)=\sigma(w^l a^{l-1} + b^l)
\end{align}
where we notice that the output of the previous layer is the input vector into the subsequent layer. All final layers in the feed-forward networks used $\sigma=\tanh$ activation functions. The output of an entire layer $a^l$ is a sequence structured as a vector that then becomes the input to the next layer. Information in this compositional model flows `forward' (hence `feed-forward'). 

When the entire set of units of a preceding layer becomes an input into each unit of the subsequent layer, we say the layer is dense. The weights are updated using backpropagation and gradient descent with respect to the applicable cost functional (description from \cite{Nielsen_2015}
 below, here $\odot$ is the Hadamard (element-wise) product), $x$ refers to each training example (batch gradient descent example below).
 
\begin{algorithm}[H]
\SetAlgoLined
 Input: Set $x = a^0$\\
 Feed-forward: For $m$ layers, for $l=2,...,m$ calculate: \\
 \qquad $z^{x,l} = w^l a^{x,l-1}+b^l$\\
 \qquad $a^{x,l} = \sigma(z^{x,l})$\\
 \qquad $\sigma=\tanh$ for $l=m$ \\
 Output layer $(L=m)$ error $\delta^{x,l}$: \\
 \qquad $\delta^{x,l} = ((w^{l+1})^T \delta^{x,l+1})
  \odot \sigma'(z^{x,l})$\\
  \qquad $\sigma'=\frac{\partial a_k^{x,L}}{\partial z^{x,L}_k}$\\
  \qquad $k$ runs over neurons in layer $L$\\
 Backpropagation: for layers $l=L-1,L-2,...,2$, calculate:\\
 \qquad $\delta^{x,L} = \nabla_a C_x \odot \sigma'(z^{x,L})$\\
 Gradient: cost function  gradient given by: \\
 
 \qquad $\frac{\partial c}{(\partial w_{jk}^{x,l}} = a^{x,l-1}_{k}\delta^{x,l}_j$ and $\frac{\partial C}{\partial b^{x,l}_j} = \delta^l_j$\\
 Update weights: for each layer $l=L,L-1,...,2$ update:\\
 \qquad $w^l \to
  w^l-\frac{\eta}{m} \sum_x \delta^{x,l} (a^{x,l-1})^T$\\
  \qquad $b^l \to b^l-\frac{\eta}{m}
  \sum_x \delta^{x,l}$
  \qquad 
 \caption{Stochastic gradient descent and backpropagation (batch) \cite{Nielsen_2015}}
\end{algorithm}

\subsection{LSTMs and GRUs}
Long-Short Term Memory networks and Gated Recurrent Units are a prevalent form of recurrent neural network (RNN). RNNs are networks tailored to modelling sequential data, such as time-series data, or data such as sequences of control amplitudes $(c_j)$ \cite{Goodfellow_Bengio_Courville_2016}. For RNNs, for each time-step $t$, there is an input $x_t$ (such as $c_t$), an output $y_t$ and hidden-layer output $h_t$. The key intuitive idea behind RNNs is that $h_t$ of the network itself becomes an input into hidden layers for the immediately next time-step $t+1$.  LSTMs advance upon this concept by enabling the output of hidden layers to influence not just the immediately succeeding time-step $t+1$, but also potentially activation functions at later time steps. In this sense LSTMs allow information about previous hidden layers (or states) to function as `memory' that is carried forward.

One of the challenges regarding RNNs is the saturation of networks where new inputs to an activation function fail to contribute significantly to its output. Intuitively too much information is saturating the model, so additional information does not lead to material updates (manifest, for example in flatlining loss, as seen in some examples above). A way to overcome this problem of saturation includes to stochastically `forget' certain information in order to make room for additional information, as manifest in the forget gate of an LSTM, distinct from the update gate. GRUs by contrast seek to incorporate the output of hidden layers and updates into subsequent hidden layers as detailed below. Their popularity is often owing to their improved speedup over LSTMs for a variety of contexts.

The reset gate combines the input $x_t$ at time $t$ with the previous time-step hidden state $h_{t-1}$ to define a reset output $r_t$ \cite{cho-etal-2014-properties}:
\begin{align*}
    r_t = \sigma(w_r x_t + u_r h_{t-1} + b_r)
\end{align*}
where $w_r, u_r$ are updatable weight matrices and $b_r$ is an applicable bias, with $\sigma$ an activation function (in our models, the tanh function to produce control amplitudes $(c_j)$ but usually the sigmoid function). The update gate remains:
\begin{align*}
    z_t = \sigma(w_z x_t + u_z h_{t-1} + b_z)
\end{align*}
This update gate is the output of the unit at time $t$. However, in order to output $h_t$, an intermediate hidden layer state is calculated:
\begin{align*}
    \tilde{h}_t = \tanh(w_h x_t + u_h(r_t \odot h_{t-1}) + b_h)
\end{align*}
where we see the $(r_t \odot h_{t-1})$ term incorporates the influence of the reset gate and previous hidden layer $h_{t-1}$ into the estimate. The final hidden layer output is then calculated by combining the Hadamard products of the update gate and previous hidden state together with the intermediate hidden state:
\begin{align*}
    h_t = z_t \odot h_{t-1} + (1-z_t) \odot \tilde{h}_t
\end{align*}
which is the ultimate output at time $t$. The incorporation of $h_{t-1}$ in this way allows influence of prior information in the sequence to influence future outputs, improving the correlation between outputs such as controls.


\end{document}